\documentclass{IET}
\usepackage{array}
\usepackage{subfigure}
\usepackage[subfigure]{graphfig}
\usepackage[linesnumbered]{algorithm2e}
\usepackage{stfloats}
\usepackage{multirow}
\usepackage{enumitem}
\usepackage{bm}
\usepackage{mathtools}
\usepackage{amsmath}
\usepackage{amssymb}   % mathematitions & symbolics
\usepackage{latexsym}   %% symbolics
\usepackage{graphicx}
\usepackage{empheq}
\usepackage{glossaries}
\newglossaryentry{ml}{name={ML},description={Machine learning}}
\newglossaryentry{svms}{name={SVMs},description={Support Vector Machines}}
\newglossaryentry{pca}{name={PCA},description={Principal Component Analysis}}
\newglossaryentry{pls}{name={PLS},description={Partial Least Squares}}
\newglossaryentry{fpm}{name={FPM},description={First Principal Model}}

\newglossaryentry{ann}{name={ANN},description={Artificial Neural Networks}}
\newglossaryentry{fnn}{name={FNNs},description={Feedforward Neural Networks}}
\newglossaryentry{rbfn}{name={RBFN},description={Radial Basis Function Networks}}
\newglossaryentry{rnn}{name={RNN},description={Recurrent Neural Networks}}
\newglossaryentry{fm}{name={FMs},description={Fuzzy Models}}
\newglossaryentry{ts}{name={T-S},description={Takagi-Sugeno}}
\newglossaryentry{fnn-1}{name={FNN},description={Fuzzy Neural Network}}

\newglossaryentry{pso}{name={PSO},description={Particle Swarm Optimization}}
\newglossaryentry{ga}{name={GA},description={Genetic Algorithm}}
\newglossaryentry{cg}{name={CG},description={Conjugate Gradient}}
\newglossaryentry{ll}{name={LL},description={Log-Likelihood}}
\newglossaryentry{map}{name={MAP},description={Maximizing A Posterior}}
\newglossaryentry{mc}{name={MC},description={Monte Carlo}}
\newglossaryentry{mcmc}{name={MCMC},description={Markov Chain Monte Carlo}}
\newglossaryentry{sor}{name={SoR},description={Subset of Regressors}}
\newglossaryentry{ivm}{name={IVM},description={Informative Vector Machine}}

\newglossaryentry{gp}{name={GP},description={Gaussian Process}}
\newglossaryentry{dgp}{name={DGP},description={Dependent Gaussian Process}}
\newglossaryentry{cgp}{name={CGP},description={Convolved Gaussian Process}}
\newglossaryentry{igp}{name={IGP},description={Independent Gaussian Process}}
\newglossaryentry{gprn}{name={GPRN},description={Gaussian Process Regression Networks}}
\newglossaryentry{lvm}{name={GP-LVM},description={Gaussian Latent Variable Model}}
\newglossaryentry{lgp}{name={LGP},description={Local Gaussian Process}}

\newglossaryentry{mimo}{name={MIMO},description={Multiple-Input Multiple-Output}}
\newglossaryentry{miso}{name={MISO},description={Multiple-Input Single-Output}}
\newglossaryentry{siso}{name={SISO},description={Single-Input Single-Output}}
\newglossaryentry{nlti}{name={NLTI},description={Nonlinear Time-Invariant}}
\newglossaryentry{nltv}{name={NLTV},description={Nonlinear Time-Varying}}
\newglossaryentry{lti}{name={LTI},description={Linear Time-Invariant}}
\newglossaryentry{ltv}{name={LTV},description={Linear Time-Varying}}

\newglossaryentry{cep}{name={CEP},description={Certainty Equivalence Principle}}
\newglossaryentry{uav}{name={UAV},description={Unmanned Aerial Vehicle}}
\newglossaryentry{gmv}{name={GMV},description={Generalized Minimum Variance}}

\newglossaryentry{cp}{name={CP},description={Convolution Process}}
\newglossaryentry{lmc}{name={LMC},description={Linear Model of Coregionalization}}
\newglossaryentry{idc}{name={IDC},description={Inverse Dynamics Control}}
\newglossaryentry{imc}{name={IMC},description={Internal Model Control}}
\newglossaryentry{mpc}{name={MPC},description={Model Predictive Control}}
\newglossaryentry{nmpc}{name={NMPC},description={ Nonlinear Model Predictive Control}}
\newglossaryentry{lmpc}{name={LMPC},description={ Linear Model Predictive Control}}
\newglossaryentry{smpc}{name={SMPC},description={ Stochastic Model Predictive Control}}
\newglossaryentry{rmpc}{name={RMPC},description={ Robust Model Predictive Control}}   
	
\newglossaryentry{mrac}{name={MRAC},description={Model References Adaptive Control}}
\newglossaryentry{dmc}{name={DMC},description={Dynamic Matrix Control}}
\newglossaryentry{pfc}{name={PFC},description={Predictive Functional Control}}
\newglossaryentry{gpc}{name={GPC},description={Generalized Predictive Control}}
\newglossaryentry{mse}{name={MSE},description={Mean Squared Error}}
\newglossaryentry{mae}{name={MAE},description={Mean Absolute Error}}
\newglossaryentry{se}{name={SE},description={Standard Error}}
\newglossaryentry{smse}{name={SMSE},description={Standardized Mean Squared Error}}
	
\newglossaryentry{te}{name={TE},description={Tennessee Eastman}}
\newglossaryentry{vtol}{name={VTOL},description={Vertical Take Off and Landing}}
\newglossaryentry{dof}{name={DOF},description={Degree-of-Freedom}}
\newglossaryentry{pid}{name={PID},description={Proportional-Integral-Derivative}}
\newglossaryentry{lqr}{name={LQR},description={Linear-Quadratic Regulator}}
\newglossaryentry{mpqp}{name={mp-QP},description={Multi-Parametric Quadratic Programs}}
\newglossaryentry{mle}{name={MLE},description={Maximum Likelihood Estimation}}
\newglossaryentry{ls}{name={LS},description={Least Square}}
\newglossaryentry{ppd}{name={PPD},description={Pseudo-Partial Derivative}}
\newglossaryentry{mfac}{name={MFAC},description={Model-Free Adaptive Control}}
\newglossaryentry{cfdl}{name={CFDL},description={Compact Form Dynamic Linearization}}
\newglossaryentry{pfdl}{name={PFDL},description={Partial Form Dynamic Linearization}}
\newglossaryentry{ddc}{name={DDC},description={Data Driven Control}}
\newglossaryentry{dp}{name={DP},description={Dynamic Programming}}
\newglossaryentry{adp}{name={ADP},description={Approximate Dynamic Programming}}
\newglossaryentry{mdp}{name={MDP},description={Markovian Decision Process}}
\newglossaryentry{pilco}{name={PILCO},description={ Probabilistic Inference for Learning Control}}
\newglossaryentry{lp}{name={LP},description={ Linear Programming}}
\newglossaryentry{nlp}{name={NLP},description={ Nonlinear Programming}} 
\newglossaryentry{kkt}{name={KKT},description={ Karush-Kahn-Tucker}}
\newglossaryentry{qp}{name={QP},description={ Quadratic Programming}}
\newglossaryentry{sqp}{name={SQP},description={ Sequential Quadratic Programming}}
\newglossaryentry{fpsqp}{name={FP-SQP},description={ Feasibility-Perturbed Sequential Quadratic Programming}}
\newglossaryentry{mfcq}{name={MFCQ},description={ Mangasarian-Fromovitz Constraint Qualification}}
\newglossaryentry{licq}{name={LICQ},description={ Linear Independence Constraint Qualification}}
\newglossaryentry{iae}{name={IAE},description={ Integral Absolute Error}}
\newglossaryentry{bfgs}{name={BFGS},description={ Broyden-Fletcher-Goldfarb-Shanno}}
\everymath{\displaystyle}

\newcommand{\singlefigWidth}{0.475}

\newcommand{\tripfigWidth}{0.3}

\begin{document}

\title{Gaussian Process Model Predictive Control of  Unknown Nonlinear Systems}

\author[1,*]{Gang~Cao}
\affil{School of Engineering and Advanced Technology, Massey University, Auckland, New Zealand.}

\author[2]{Edmund M-K~Lai}
\affil{Department of Information Technology and Software Engineering, Auckland University of Technology, Auckland, New Zealand.}

\author[1]{Fakhrul~Alam}

\affil[*]{g.cao@massey.ac.nz}

%\author{Gang~Cao, Edmund M-K~Lai and Fakhrul~Alam%
%\thanks{Manuscript to be submitted to IEEE Transactions on
%	Control Systems Technology.}%
%\thanks{G. Cao and F. Alam are with the School of Engineering and Advanced Technology, Massey University, Auckland, New Zealand. Email: \{g.cao,f.alam\}@massey.ac.nz}%
%\thanks{E. M-K. Lai is with the Department of Information Technology and Software Engineering, Auckland University of Technology, Auckland, New Zealand. Email: edmund.lai@aut.ac.nz}
%}

\abstract{\gls{mpc} of an unknown system that is modelled by \gls{gp} techniques is studied in this paper.
Using~\gls{gp}, the variances computed during the modelling and inference processes
allow us to take model uncertainty into account.
The main issue in using \gls{mpc} to control systems modelled by \gls{gp} is the propagation of
such uncertainties within the control horizon.
In this paper, two approaches to solve this problem, called GPMPC1 and GPMPC2, are proposed.
With GPMPC1, the original~\gls{smpc} problem is relaxed to a deterministic nonlinear~\gls{mpc} 
based on a basic linearized \gls{gp} local model.
The resulting optimization problem, though non-convex, can be solved by the \gls{sqp}.
By incorporating the model variance into the state vector, an extended local model is derived.
This model allows us to relax the non-convex~\gls{mpc} problem to a convex one which can be solved
by an active-set method efficiently.
The performance of both approaches is demonstrated by applying them to two trajectory tracking problems.
Results show that both GPMPC1 and GPMPC2 produce effective controls
but GPMPC2 is much more efficient computationally.}

\maketitle

%resets all acronyms to not used. Useful after the abstract to redefine all acronyms in the introduction.
\glsresetall
\pagestyle{empty}    % remove page numbers

\newcommand*{\meanvalue}{\bm{\mu}}
\newcommand*{\varvalue}{\bm{\Sigma}}
\newcommand*{\hyperparas}{\bm{\theta}}

\section{Introduction}
\label{sec:introduction}
\thispagestyle{empty} 

\gls{mpc}, also known as receding horizon control,
is a class of computer control algorithms that predicts future responses of a plant based on its system model,
and computes optimized control inputs by repeatedly solving a finite horizon optimization problem~\cite{A-Mayne-ConstrainedMPC-StabilityOptimality-2000}.
The advantages of \gls{mpc} mainly lie in its conceptual simplicity for multiple variable problems,
and its ability to handle input and output ``hard-constraints" that are commonly encountered in practice 
but are not well addressed by other control methods. 
It has been applied to many different types of control problems~\cite{A-Qin-SurveyMPC-2003,A-Mayne-ServeyMPC-2014}.

The performance of \gls{mpc} is highly dependent on the accuracy of the system model that describes its dynamics.
Traditionally, these models are derived mathematically.
More recently, data-driven modelling approaches based on computational intelligence and machine learning techniques 
are becoming popular~\cite{A-Dimitri-SurveyDataDrivenModel-2008, B-Nelles-NonlinearSystemIdentify-2013}.
This approach is especially suitable for complex and highly nonlinear systems where 
complete knowledge of the system dynamics is seldom available,
giving rise to unmodelled dynamics or model uncertainty.
From the \gls{mpc} perspective, attempts to address the issue of model uncertainty has been made through \gls{rmpc} schemes
such as open-loop ``min-max"~\gls{mpc}~\cite{A-Alamo-OpenLoopMinMaxMPC-2005}, closed-loop ``min-max"~\gls{mpc}~\cite{A-Limon-ISSMinMaxMPC-2006} and tube-based~\gls{mpc}~\cite{A-Langson-TubeMPC-2004}.
``Min-max"~\gls{mpc} is conceptually simple.
However, its control laws are computed based on worst-case scenarios and are therefore considered too conservative.
Tube-based~\gls{mpc} overcomes this problem by combining
a conventional~\gls{mpc} for the nominal system and a local feedback control law that steers the states of the unknown system to the inside of a ``tube" centered on the nominal trajectory~\cite{A-Zhang-SwitchedMPC-2016}.
This ``tube", which relates to the uncertainty bounds, must be carefully defined.
Otherwise, there may not be a feasible solution.
The major problem with \gls{rmpc} is that model uncertainties are assumed to be 
deterministic even though they are typically stochastic.

\gls{smpc} is an alternative where
model uncertainties are assumed to be stochastic with an underlying probability distribution~\cite{A-Schwarm-ChanceConstrainedMPC-1999, IP-Bernardini-ScenarioSMPC-2009, A-Cannon-ProbConstrainsSMPC-2011,IP-Mesbah-StochasticMPC-2014}.
Control laws are computed by solving a stochastic optimization problem.
Furthermore, since the state or output constraints are also probabilistic,
they can be satisfied with a predefined level of confidence.
This effectively alleviates the conservatism of ``min-max" \gls{mpc}.
Furthermore, it is possible to trade-off control performance with robustness against model uncertainties 
by adjusting these probabilistic constraints.
A key problem with \gls{smpc} is the propagation of uncertainties over a finite prediction horizon.
The most common solution is to use sampling-based \gls{mc} simulation techniques.
However, they are computationally demanding.
More recently, a technique known as polynomial chaos expansions has been proposed to 
lighten the computation burden~\cite{IP-Fagiano-NMPCusingPC-2012,IP-Mesbah-StochasticMPC-2014}.

A model known as \gls{gp} has become very useful in statistical modelling \cite{B-GPMLbook-2006}. 
The \gls{gp} variances which are computed as part of the modelling process provide a useful
indication of the accuracy of the model. These variances can also be propagated in multiple-step ahead predictions.
The hyperparameters of these models are learnt from data by maximizing the log-likelihood function.
This optimization problem is unconstrained, nonlinear and non-convex optimization.
It is typically solved by \gls{cg}~\cite{B-GPMLbook-2006} or by \gls{pso} techniques~\cite{IP-Zhu-Particle-2010,IP-Petelin-EvolvingGP-2011,IP-Gang-2014a}.

A~\gls{gp} based \gls{mpc} scheme was first introduced in~\cite{IP-Kocijan-MPC-GP-2004}.
Subsequently, an \gls{smpc} scheme using \gls{gp} was proposed in~\cite{IP-Grancharova-ExplicitMPC-2007}.
Even though \gls{gp} is a probabilistic model, the cost functions used in these papers 
are deterministic.
Consequently, the variances could only be treated as slack variables of the state constraints.
This indirect way of handling GP variances leads to a nonlinear optimization problem that
is very computationally demanding to solve.
More recently, in \cite{A-Klenske-GPMPC4Periodic-2015,IP-GPMPC4LTV-Gang-2016a,IP-GPMPC4Quad-Gang-2016b},
 variances are included in the cost function and can be directly handled
in the optimization process.
However, only unconstrained \gls{mpc} have been considered.

In this paper, two new \gls{gp} based \gls{mpc} approaches,
referred to as GPMPC1 and GPMPC2, are proposed for the control of unknown nonlinear dynamical 
systems with input and state constraints.
The GPMPC1 approach is similar to those in~\cite{IP-Kocijan-MPC-GP-2004,IP-Grancharova-ExplicitMPC-2007}
in the sense that the \gls{gp} variances are considered as a slack variable in the state constraints.
The main difference is that the resulting non-convex optimization problem is solved by using
a~\gls{sqp} based method together with a linearized \gls{gp} model which is called 
the basic~\gls{gp} based local model in this paper.
The constrained stochastic problem is then relaxed to a deterministic one 
by specifying the confidence level.
With GPMPC2, the nonlinear~\gls{mpc} problem is reformulated to a convex optimization problem.
In contrast with earlier methods,
\gls{gp} variances are directly included in the cost function of the optimization.
The solution method makes use of a modified version of the basic local model
which includes the variance in the modified state variable of the system.
The resulting~\gls{mpc} problem is efficiently solved by using an active-set method.
The effectiveness of these approaches are demonstrated
by applying them to 
two trajectory tracking problems.

The rest of this paper is organized as follows.
Section~\ref{sec:gps} introduces the modelling of the unknown nonlinear system by using~\gls{gp} models.
The basic and extended~\gls{gp} based local dynamical models are presented in the Section~\ref{sec:gplocalmodels}.
In Section~\ref{sec:gpsmpc},
the proposed GPMPC1 and GPMPC2 are presented for the general trajectory tracking problem of the unknown nonlinear system.
In addition, the feasibility and stability of the proposed algorithms are also discussed.
The simulation results are next reported to demonstrate the performance of the proposed algorithms in Section~\ref{sec:simulation}.
Finally, Section~\ref{sec:conclude} draws the conclusions.

\section{Unknown system modelling using GP}
\label{sec:gps}

Consider a discrete-time nonlinear dynamical system described by the following general form:
\begin{equation}
\mathbf{x}_{k+1} = f(\mathbf{x}_k, \mathbf{u}_k) +\mathbf{w}_k
\label{eqn:nltvsys}
\end{equation}
where $f:\mathbb{R}^{n}\times\mathbb{R}^{m}\rightarrow\mathbb{R}^{n}$ is a nonlinear function,
$\mathbf{w}\in~\mathbb{R}^n$ represents additive external disturbances,
$\mathbf{x} \in \mathbb{R}^n$ denotes the state vector, and
$\mathbf{u}\in\mathbb{R}^m$ are control signals.
In this paper, we assume that $f$ is totally unknown but can be represented by a \gls{gp} model.
The uncertainty of a \gls{gp} model can be measured by the \gls{gp} variances.
Therefore, a disturbance observer will not be required.
The hyperparameters of a \gls{gp} model is learnt from a
set of training data consisting of inputs to the system and the system's response as target.

To model a system given by (\ref{eqn:nltvsys}),
a natural choice of the model inputs and their targets are
the state-control tuple $\tilde{\mathbf{x}}_k = (\mathbf{x}_k,\mathbf{u}_k)\in\mathbb{R}^{n+m}$ 
and the next state $\mathbf{x}_{k+1}$ respectively.
Let $\Delta\mathbf{x}_k=\mathbf{x}_{k+1}-\mathbf{x}_{k}\in\mathbb{R}^n$.
In practice, the variation between $\Delta\mathbf{x}_{k+1}$ and $\Delta\mathbf{x}_k$ 
is much less the variation between $\mathbf{x}_{k+1}$ and $\mathbf{x}_k$, for all $k$.
Therefore it is more advantageous to use $\Delta\mathbf{x}_k$ as the model target instead~\cite{PHD-Deisenroth-EfficientRLusingGP-2010}.
This will be assumed in the rest of this paper.

\subsection{GP Modelling}
\label{subsec:gpmodelling}

A~\gls{gp} model is completely specified by its mean and covariance function~\cite{B-GPMLbook-2006}.
Assuming that the mean of the model input $\tilde{\mathbf{x}}_k$ is zero, the squared exponential covariance is given by $\mathbf{K}(\tilde{\mathbf{x}}_i,\tilde{\mathbf{x}}_j)=\sigma_s^2 \exp(-\frac{1}{2}(\tilde{\mathbf{x}}_i-\tilde{\mathbf{x}}_j)^T\bm{\varLambda}(\tilde{\mathbf{x}}_i-\tilde{\mathbf{x}}_j))+\sigma_n^2$. 
The parameters
$\sigma_s^2,\sigma_n^2$ and the entries of matrix $\bm{\varLambda}$ are referred to as the 
hyperparameters $\hyperparas$ of a~\gls{gp} model.
Given $D$ training inputs $\tilde{\mathbf{X}}=[\tilde{\mathbf{x}}_1,\cdots,\tilde{\mathbf{x}}_D]$ and their corresponding training targets $\mathbf{y}=[\Delta\mathbf{x}_1,\cdots,\Delta\mathbf{x}_D]^T$,
the joint distribution between $\mathbf{y}$ and a test target $\Delta\mathbf{x}^*_k$ for training input $\tilde{\mathbf{x}}_k^*$ 
is assumed to follow a Gaussian distribution. That is,
\begin{equation}
p\Bigg(\begin{array}{c} \mathbf{y}\\ \Delta\mathbf{x}^*_k
\end{array}\Bigg)
\sim \mathcal{N}\Bigg( 
0, \begin{array}{cc}
\mathbf{K}(\tilde{\mathbf{X}}, \tilde{\mathbf{X}})+\sigma_n\mathbf{I} & \mathbf{K}(\tilde{\mathbf{X}}, \tilde{\mathbf{x}}^*_k)\\
\mathbf{K}(\tilde{\mathbf{x}}^*_k, \tilde{\mathbf{X}}) & \mathbf{K}(\tilde{\mathbf{x}}^*_k, \tilde{\mathbf{x}}^*_k)
\end{array}
\Bigg)
\end{equation}
In addition,
the posterior distribution over the observations can be obtained by restricting the joint distribution to only contain those targets that agree with the observations~\cite{B-GPMLbook-2006}.
This is achieved by conditioning the joint distribution on the observations,
and results in the predictive mean and variance function as follows:
\begin{subequations}\label{eqn:meanvar}
\begin{align}
m(\tilde{\mathbf{x}}^*_k)=\textit{E}_f[\Delta\mathbf{x}^*_k]&=\mathbf{K}(\tilde{\mathbf{x}}^*_k, \tilde{\mathbf{X}})\mathbf{K}_\sigma^{-1}\mathbf{y}\\
\sigma^2(\tilde{\mathbf{x}}^*_k)=\textit{Var}_f[\Delta\mathbf{x}^*_k] &=
\mathbf{K}(\tilde{\mathbf{x}}^*_k, \tilde{\mathbf{x}}^*_k)\\
\nonumber&-\mathbf{K}(\tilde{\mathbf{x}}^*_k, \tilde{\mathbf{X}})\mathbf{K}_\sigma^{-1}\mathbf{K}(\tilde{\mathbf{X}}, \tilde{\mathbf{x}}^*_k)
\end{align}
\end{subequations}
where $\mathbf{K}_\sigma=\mathbf{K}(\tilde{\mathbf{X}}, \tilde{\mathbf{X}})+\sigma_n\mathbf{I}$.
The state at the next sampling time also follows a Gaussian distribution. Thus,
\begin{equation}
\label{eqn:nextstate}
p(\mathbf{x}_{k+1})\sim\mathcal{N}(\meanvalue_{k+1},\varvalue_{k+1})
\end{equation} 
where
\begin{subequations}\label{eqn:statedistribute-1}
\begin{align}
\meanvalue_{k+1}&=\mathbf{x}_k+m(\tilde{\mathbf{x}}^*_k)\\
\quad \varvalue_{k+1}&=\sigma^2(\tilde{\mathbf{x}}^*_k)
\end{align}
\end{subequations}

Typically, the hyperparameters of the \gls{gp} model are learned by maximizing the log-likelihood function given by
\begin{equation}
\log p(\mathbf{y}|\tilde{\mathbf{X}},\hyperparas)=
-\frac{1}{2}\mathbf{y}^T\mathbf{K}_\sigma^{-1}\mathbf{y}
-\frac{1}{2}\log\left|\mathbf{K}_\sigma^{-1}\right| -\frac{D}{2}\log(2\pi)
\label{eqn:loglikelihood}
\end{equation}
This results in a nonlinear non-convex optimization problem that is traditionally solved by using~\gls{cg} or~\gls{bfgs} algorithms.
Recently,
\gls{pso} based algorithms that minimizes the model error instead of the log-likelihood function
have been shown in~\cite{IP-Gang-2014a} to be more efficient and effective.

\subsection{Uncertainty propagation}
\label{subsec:undertaintypropagate}
 
With the \gls{gp} model obtained, one-step-ahead predictions can be made by using (\ref{eqn:meanvar}) 
and (\ref{eqn:statedistribute-1}).
When multiple-step predictions are required,
the conventional way is to iteratively perform multiple one-step-ahead predictions using the estimated mean values.
However, this process does not take into account the uncertainties introduced by each successive prediction.
This issue has been shown to be important in time-series predictions~\cite{IP-Girard-MultiStepTimeSeriesForecasting-2003}.

The uncertainty propagation problem can be dealt with by assuming that the joint distribution of 
the training inputs is uncertain and follows a Gaussian distribution.
That is,
\begin{equation}
p(\tilde{\mathbf{x}}_k)=p\big(\mathbf{x}_k,\mathbf{u}_k\big)\sim \mathcal{N}(\tilde{\meanvalue}_k,\tilde{\varvalue}_k)
\end{equation} 
with mean and variance given by
\begin{subequations}\label{eqn:augementedstateMeanVar}
\begin{align}
\tilde{\meanvalue}_k &=\left[\meanvalue_k,\textit{E}\left[\mathbf{u}_k\right]\right]^T \\
\tilde{\varvalue}_k &=\left[\begin{array}{cc}
\varvalue_k & \textit{Cov}\left[\mathbf{x}_k,\mathbf{u}_k\right]\\
\textit{Cov}\left[\mathbf{u}_k,\mathbf{x}_k\right]&
\textit{Var}\left[\mathbf{u}_k\right]
\end{array}
\right]
\end{align}
\end{subequations}
where 
$\textit{Cov}\left[\mathbf{x}_k,\mathbf{u}_k\right] =
\textit{E}\left[\mathbf{x}_k\mathbf{u}_k\right]-\meanvalue_k\textit{E}\left[\mathbf{u}_k\right]$. 
Here, $\textit{E}\left[\mathbf{u}_k\right]$ and $\textit{Var}\left[\mathbf{u}_k\right]$ are the mean and variance of the system controls.

The exact predictive distribution of the training target could then be obtained by integrating over the training input distribution: 
\begin{equation}
p(\Delta\mathbf{x}^*_k)=\int p(f(\tilde{\mathbf{x}}^*_k)|\tilde{\mathbf{x}}^*_k) p(\tilde{\mathbf{x}}^*_k)d\tilde{\mathbf{x}}^*_k
\end{equation}
However, this integral is analytically intractable.
Numerical solutions can be obtained using Monte-Carlo simulation techniques.
In~\cite{IP-Candela-GPUncertainPropagation-2003}, a moment-matching based approach is proposed to obtain an analytical Gaussian approximation.
The mean and variance at an uncertain input can be obtained
through the laws of iterated expectations and conditional variances respectively~\cite{PHD-Deisenroth-EfficientRLusingGP-2010}.
They are given by
\begin{subequations}\label{eqn:pgpmeanvar}
\begin{align}
m(\tilde{\mathbf{x}}^*_k) & =
\textit{E}_{\tilde{\mathbf{x}}^*_k}\Big[\textit{E}_f\big[\Delta\mathbf{x}^*_k\big]\Big]\\
\sigma^2(\tilde{\mathbf{x}}^*_k) &=\textit{E}_{\tilde{\mathbf{x}}^*_k}\Big[\textit{Var}_f\big[\Delta\mathbf{x}^*_k\big]\Big]+\textit{Var}_{\tilde{\mathbf{x}}^*_k}\Big[\textit{E}_f\big[\Delta\mathbf{x}^*_k\big]\Big]
\end{align}
\end{subequations}
Equation (\ref{eqn:statedistribute-1}) then becomes
\begin{subequations}\label{eqn:statedistribute-2}
\begin{align}
\meanvalue_{k+1} =& \meanvalue_k + m(\tilde{\mathbf{x}}^*_k)\\
\varvalue_{k+1} =& \varvalue_k+ \sigma^2(\tilde{\mathbf{x}}^*_k)\\
\nonumber&+\textit{Cov}\big[\mathbf{x}_k, \Delta\mathbf{x}_k\big]+\textit{Cov}\big[\Delta\mathbf{x}_k, \mathbf{x}_k\big]
\end{align}
\end{subequations}

The computational complexity of~\gls{gp} inference using (\ref{eqn:pgpmeanvar}) is $\mathcal{O}(D^2n^2(n+m))$
which is quite high.
Hence, \gls{gp} is normally only suitable for problems with limited dimensions (under 12 as suggested by most publications) and limited size of training data.
For problems with higher dimensions,
sparse~\gls{gp} approaches~\cite{A-Sparse-Appro-2005} are often used. 

\section{GP Based Local Dynamical Models}
\label{sec:gplocalmodels}

When dealing with the control of nonlinear systems, it is common practice
to obtain local linearized models of the system around operating points.
The main purpose is to reduce the computation involved in the nonlinear control problem.
The same technique is used here for the \gls{gp} based~\gls{mpc} optimization problem.
The main difference here is that the model of the system is probabilistic rather than deterministic. 
Thus there is more than one way by which the \gls{gp} model could be linearized.

In~\cite{IP-Berkenkamp-RboustLBNMPC-2014}, a \gls{gp} based local dynamical model allows
standard robust control methods to be used on the partially unknown system directly.
Another~\gls{gp} based local dynamical model is proposed in~\cite{IP-Pan-PDDP-2014} to integrate \gls{gp} model 
with dynamic programming. 
In these two cases, the nonlinear optimization problems considered are \textit{unconstrained}.

In this section, we shall present two different~\gls{gp} based local models.
They will be applied to the \textit{constrained} nonlinear problems presented in Section~\ref{sec:gpsmpc}.

\subsection{Basic GP-based Local Model}
\label{subsec:basicLinear}
Linearization can be done based on the mean values in the \gls{gp} model.
In this case we replace the state vector $\mathbf{x}_k$ by its mean $\meanvalue_k$.
Then (\ref{eqn:nltvsys}) becomes
\begin{equation}
\meanvalue_{k+1}=\mathcal{F}(\meanvalue_k, \mathbf{u}_k)
\label{eqn:gpdynamicmodel-1}
\end{equation}
Let ($\meanvalue_k^*,\mathbf{u}^*_k$) be the operating point at which the linearized model is to be obtained.
Given that $\Delta\meanvalue_k = \meanvalue_k-\meanvalue_k^*$ and $\Delta\mathbf{u}_k = \mathbf{u}_k-\mathbf{u}_k ^*$
are small, from (\ref{eqn:gpdynamicmodel-1}), we have
\begin{subequations}\label{eqn:gplocalmodel-1}
\begin{align}
\Delta\meanvalue_{k+1} &=\frac{\partial\mathcal{F}}{\partial\meanvalue_k}\Delta\meanvalue_k+\frac{\partial\mathcal{F}}{\partial\mathbf{u}_k}\Delta\mathbf{u}_k\\
&=\frac{\partial\meanvalue_{k+1}}{\partial\meanvalue_k}\Delta\meanvalue_k+\frac{\partial\meanvalue_{k+1}}{\partial\mathbf{u}_k}\Delta\mathbf{u}_k
\end{align}
\end{subequations}
where $\frac{\partial\meanvalue_{k+1}}{\partial\meanvalue_k}$ and $\frac{\partial\meanvalue_{k+1}}{\partial\mathbf{u}_k}$ 
are the Jacobian state and input matrices respectively.
Using the chain rule, we get
\begin{subequations}
\begin{align}
\frac{\partial\meanvalue_{k+1}}{\partial\meanvalue_k}
&=\frac{\partial\meanvalue_{k+1}}{\partial\tilde{\meanvalue}_k}\frac{\partial\tilde{\meanvalue}_k}{\partial\meanvalue_k}
+\frac{\partial\meanvalue_{k+1}}{\partial\tilde{\varvalue}_k}\frac{\partial\tilde{\varvalue}_k}{\partial\meanvalue_k}\\
\frac{\partial\meanvalue_{k+1}}{\partial\mathbf{u}_k} &=\frac{\partial\meanvalue_{k+1}}{\partial\tilde{\meanvalue}_k}\frac{\partial\tilde{\meanvalue}_k}{\partial\mathbf{u}_k}
+\frac{\partial\meanvalue_{k+1}}{\partial\tilde{\varvalue}_k}\frac{\partial\tilde{\varvalue}_k}{\partial\mathbf{u}_k}
\end{align}
\end{subequations}
where
$\frac{\partial\tilde{\meanvalue}_k}{\partial\meanvalue_k}$,
$\frac{\partial\tilde{\varvalue}_k}{\partial\meanvalue_k}$, 
$\frac{\partial\tilde{\meanvalue}_k}{\partial\mathbf{u}_k}$, 
$\frac{\partial\tilde{\varvalue}_k}{\partial\mathbf{u}_k}$ 
can be easily obtained based on (\ref{eqn:augementedstateMeanVar}).
Elaborations of 
$\frac{\partial\meanvalue_{k+1}}{\partial\tilde{\meanvalue}_k}$ and
$\frac{\partial\meanvalue_{k+1}}{\partial\tilde{\varvalue}_k}$
can be found in~\cite{PHD-Deisenroth-EfficientRLusingGP-2010}.

\subsection{Extended GP-based Local Model}
\label{subsec:extendedLinear}
Model uncertainties are characterized by the variances.
However, the basic local model derived above only involves the mean values.
The extended local model aims to take into account model uncertainties.
Similar to what we have done to derive the basic model, we replace the state vector $\mathbf{x}_k$ in (\ref{eqn:nltvsys})
by 
$\mathbf{s}_k= [\meanvalue_k, \textbf{vec}(\sqrt{\varvalue_k})]^T\in \mathbb{R}^{n+n^2}$
which shall be known as the ``extended state".
Here, $\textbf{vec}(\cdot)$ denotes the vectorization of a matrix
~\footnote{$\varvalue_k$ is a real symmetric matrix therefore can be diagonalized. 
The square root of a diagonal matrix can simply be obtained by computing the square roots of diagonal entries.}.
Hence (\ref{eqn:nltvsys}) becomes
\begin{equation}
\mathbf{s}_{k+1}=\mathcal{F}'\left(\mathbf{s}_k, \mathbf{u}_k\right) 
\label{eqn:gpdynamicmodel-2}
\end{equation}
Linearizing at the operating point ($\mathbf{s}_k^*, \mathbf{u}^*_k$)
where $\mathbf{s}_k^* = [\meanvalue^*_k,\textbf{vec}(\sqrt{\varvalue^*_k})]^T$, we have
\begin{equation}
\Delta\mathbf{s}_{k+1} =\frac{\partial\mathcal{F}'}{\partial\mathbf{s}_k}\Delta\mathbf{s}_k+\frac{\partial\mathcal{F}'}{\partial\mathbf{u}_k}\Delta\mathbf{u}_k
\label{eqn:gplocalmodel-2}
\end{equation}
Here, $\Delta\mathbf{s}_k = \mathbf{s}_k-\mathbf{s}_k^*$ and $\Delta\mathbf{u}_k = \mathbf{u}_k-\mathbf{u}_k ^*$.
The Jacobian matrices are
\begin{subequations}\label{eqn:extendgpJacobian}
\begin{align}
\frac{\partial\mathcal{F}'}{\partial\mathbf{s}_k}&=
\left[
\begin{array}{cc}
\frac{\partial\meanvalue_{k+1}}{\partial\meanvalue_k} & \frac{\partial\meanvalue_{k+1}}{\partial\sqrt{\varvalue_k}}\\
\frac{\partial\sqrt{\varvalue_{k+1}}}{\partial\meanvalue_k} &
\frac{\partial\sqrt{\varvalue_{k+1}}}{\partial\sqrt{\varvalue_k}}
\end{array}
 \right]\in \mathbb{R}^{(n+n^2)\times(n+n^2)} \\
\frac{\partial\mathcal{F}'}{\partial\mathbf{u}_k}&=
\left[
\begin{array}{c}
\frac{\partial\meanvalue_{k+1}}{\partial\mathbf{u}_k}\\
\frac{\partial\sqrt{\varvalue_{k+1}}}{\partial\mathbf{u}_k}
\end{array}
 \right] \in \mathbb{R}^{(n+n^2)\times m}
\end{align}
\end{subequations}
with the entries given by
\begin{subequations}\label{eqn:computationextendgpJacobian}
\begin{align}
\frac{\partial\meanvalue_{k+1}}{\partial\sqrt{\varvalue_k}}&=
\frac{\partial\meanvalue_{k+1}}{\partial\varvalue_k}
\frac{\partial\varvalue_k}{\partial\sqrt{\varvalue_k}}\\
\frac{\partial\sqrt{\varvalue_{k+1}}}{\partial\meanvalue_k}&=
\frac{\partial\sqrt{\varvalue_{k+1}}}{\partial\varvalue_{k+1}}
\frac{\partial\varvalue_{k+1}}{\partial\meanvalue_k}\\
\frac{\partial\sqrt{\varvalue_{k+1}}}{\partial\sqrt{\varvalue_k}}&=
\frac{\partial\sqrt{\varvalue_{k+1}}}{\partial\varvalue_{k+1}}
\frac{\partial\varvalue_{k+1}}{\partial\varvalue_k}
\frac{\partial\varvalue_k}{\partial\sqrt{\varvalue_k}}\\
\frac{\partial\sqrt{\varvalue_{k+1}}}{\partial\mathbf{u}_k}&=
\frac{\partial\sqrt{\varvalue_{k+1}}}{\partial\varvalue_{k+1}}
\frac{\partial\varvalue_{k+1}}{\partial\mathbf{u}_k}
\end{align}
\end{subequations}
Since $\frac{\partial\sqrt{\varvalue_k}}{\partial\varvalue_k}=\frac{1}{2\sqrt{\varvalue_k}}$ and $\frac{\partial\sqrt{\varvalue_{k+1}}}{\partial\varvalue_{k+1}}=\frac{1}{2\sqrt{\varvalue_{k+1}}}$,
they can be expressed as
\begin{subequations}
\begin{align}
\frac{\partial\meanvalue_{k+1}}{\partial\varvalue_k} &=\frac{\partial\meanvalue_{k+1}}{\partial\tilde{\meanvalue}_k}\frac{\partial\tilde{\meanvalue}_k}{\partial\varvalue_k}
+\frac{\partial\meanvalue_{k+1}}{\partial\tilde{\varvalue}_k}\frac{\partial\tilde{\varvalue}_k}{\partial\varvalue_k}\\
\frac{\partial\varvalue_{k+1}}{\partial\meanvalue_k} &=\frac{\partial\varvalue_{k+1}}{\partial\tilde{\meanvalue}_k}\frac{\partial\tilde{\meanvalue}_k}{\partial\meanvalue_k}
+\frac{\partial\varvalue_{k+1}}{\partial\tilde{\varvalue}_k}\frac{\partial\tilde{\varvalue}_k}{\partial\meanvalue_k}\\
\frac{\partial\varvalue_{k+1}}{\partial\varvalue_k} &=\frac{\partial\varvalue_{k+1}}{\partial\tilde{\meanvalue}_k}\frac{\partial\tilde{\meanvalue}_k}{\partial\varvalue_k}
+\frac{\partial\varvalue_{k+1}}{\partial\tilde{\varvalue}_k}\frac{\partial\tilde{\varvalue}_k}{\partial\varvalue_k}\\
\frac{\partial\varvalue_{k+1}}{\partial\mathbf{u}_k} &=\frac{\partial\varvalue_{k+1}}{\partial\tilde{\meanvalue}_k}\frac{\partial\tilde{\meanvalue}_k}{\partial\mathbf{u}_k}
+\frac{\partial\varvalue_{k+1}}{\partial\tilde{\varvalue}_k}\frac{\partial\tilde{\varvalue}_k}{\partial\mathbf{u}_k}
\end{align}
\end{subequations}
$\frac{\partial\tilde{\meanvalue}_k}{\partial\varvalue_k}$ and 
$\frac{\partial\tilde{\varvalue}_k}{\partial\varvalue_k}$ can be easily obtained based on (\ref{eqn:augementedstateMeanVar}).
Elaborations of
$\frac{\partial\varvalue_{k+1}}{\partial\tilde{\meanvalue}_k}$ and
$\frac{\partial\varvalue_{k+1}}{\partial\tilde{\varvalue}_k}$
can be found in~\cite{PHD-Deisenroth-EfficientRLusingGP-2010}.

\section{Model Predictive Control based on GP}
\label{sec:gpsmpc}

A discrete-time nonlinear dynamical system defined by (\ref{eqn:nltvsys})
is required to track a trajectory given by  $\{\mathbf{r}_k\}$ for $k=1,2,\cdots$.
Using \gls{mpc} with a prediction horizon $H\geq 1$,
the optimal control sequence
can be obtained by solving the following problem:
\begin{subequations}\label{eqn:optproblem1}
\begin{align}
\mathbf{V}_k^*&=\min_{\mathbf{u}(\cdot)}\mathcal{J}(\mathbf{x}_k,\mathbf{u}_{k-1}, \mathbf{r}_k)\\
\mbox{s.t.}\; & \mathbf{x}_{k+i|k}=f(\mathbf{x}_{k+i-1|k}, \mathbf{u}_{k+i-1})\\
\label{eqn:constraints-x}&\mathbf{x}_{\text{min}} \leq \mathbf{x}_{k+i|k} \leq \mathbf{x}_{\text{max}}\\
\label{eqn:constraints-u}&\mathbf{u}_{\text{min}} \leq\mathbf{u}_{k+i-1} \leq \mathbf{u}_{\text{max}}\\
\nonumber& i=1,\cdots, H
\end{align}
\end{subequations}
where only the first control action $\mathbf{u}_k$ of the resulting control sequence $\mathbf{u}(\cdot)=[\mathbf{u}_k,\cdots,\mathbf{u}_{k+H-1}]^T$ is applied to the system at time $k$.
$\mathbf{x}_{\text{min}}\leq\mathbf{x}_{\text{max}}$ and $\mathbf{u}_{\text{min}}\leq\mathbf{u}_{\text{max}}$ are the upper and lower bounds of the system states and control inputs, respectively.

In the rest of this paper, the cost function $\mathcal{J}(\mathbf{x}_k,\mathbf{u}_{k-1},\mathbf{r}_k)$ shall be rewritten as $\mathcal{J}(\mathbf{x}_k,\mathbf{u}_{k-1})$ for brevity.
The quadratic cost function given by
\begin{equation}
\begin{aligned}
\mathcal{J}(\mathbf{x}_k,\mathbf{u}_{k-1})=\sum_{i=1}^{H}\Big\{\big\|\mathbf{x}_{k+i}-\mathbf{r}_{k+i}\big\|^2_{\mathbf{Q}}+\big\|\mathbf{u}_{k+i-1}\big\|^2_{\mathbf{R}}\Big\}
\end{aligned}
\label{eqn:costfunction1}
\end{equation}
will be used. 
Here, $\big\|\cdot\big\|_{\mathbf{Q}}$ and $\big\|\cdot\big\|_{\mathbf{R}}$ denote the two $2$-norms weighted by positive definite matrices $\mathbf{Q} \in \mathbb{R}^{n\times n}$ and $\mathbf{R} \in \mathbb{R}^{m\times m}$ respectively.
The control horizon will be assumed to be equal to the prediction horizon.

\subsection{GPMPC1}
\label{sec:gpmpc1}

\subsubsection{Problem Reformulation}
We assume that the system function $f(\cdot)$ is unknown and it is replaced by a \gls{gp} model.
%with the consideration of the uncertainty propagation given in Section~\ref{sec:gps},
%the multi-step states over the control horizon $H$ can be updated by using (\ref{eqn:statedistribute-2}).
Consequently, problem (\ref{eqn:optproblem1}) becomes a stochastic one~\cite{A-Grancharova-ExplicitMPC-2008}: 
\begin{subequations}\label{eqn:optproblem2}
\begin{align}
\mathbf{V}_k^\ast =& 
\min_{\mathbf{u}(\cdot)}\textit{E}\big[ \mathcal{J}(\mathbf{x}_k,\mathbf{u}_{k-1})\big]\\
\mbox{s.t.}\quad & p(\mathbf{x}_{k+1}|\mathbf{x}_k)\sim\mathcal{N}(\meanvalue_{k+1},\varvalue_{k+1})\\
&\mathbf{u}_{\text{min}} \leq\mathbf{u}_{k+i-1} \leq \mathbf{u}_{\text{max}}\\
& p\big\{\mathbf{x}_{k+i|k}\geq \mathbf{x}_{\text{min}}\big\} \geq \eta \label{subeqn:chance1}\\
& p\big\{\mathbf{x}_{k+i|k}\leq \mathbf{x}_{\text{max}}\big\} \geq \eta \label{subeqn:chance2}
\end{align}
\end{subequations}
where $\eta$ denotes a confidence level. 
For $\eta = 0.95$, the chance constraints (\ref{subeqn:chance1})  and (\ref{subeqn:chance2})
are equivalent to
\begin{subequations}\label{eqn:deterministicConstraints}
\begin{align}
\meanvalue_{k+i}-2\varvalue_{k+i} &\geq \mathbf{x}_{\text{min}}\\
\meanvalue_{k+i}+2\varvalue_{k+i} &\leq \mathbf{x}_{\text{max}}
\end{align}
\end{subequations}
Using (\ref{eqn:costfunction1}) as the cost function, we get
\begin{equation}
\begin{aligned}
 &\textit{E}\big[\mathcal{J}(\mathbf{x}_k,\mathbf{u}_{k-1})\big]\\ 
 &=\textit{E}\Big[
 \sum_{i=1}^{H}\Big\{\big\|\mathbf{x}_{k+i}-\mathbf{r}_{k+i}\big\|^2_{\mathbf{Q}}+\big\|\mathbf{u}_{k+i-1}\big\|^2_{\mathbf{R}}\} \Big] \\
 &= \sum_{i=1}^{H}\textit{E}\Big[ \big\|\mathbf{x}_{k+i}-\mathbf{r}_{k+i}\big\|^2_{\mathbf{Q}}+\big\|\mathbf{u}_{k+i-1}\big\|^2_{\mathbf{R}}\Big] \\
 &= \sum_{i=1}^{H}\bigg\{\textit{E}\Big[ \big\|\mathbf{x}_{k+i}-\mathbf{r}_{k+i}\big\|^2_{\mathbf{Q}}\Big]+\textit{E}\Big[\big\|\mathbf{u}_{k+i-1}\big\|^2_{\mathbf{R}}\Big]\bigg\}
\end{aligned}
\label{eqn:costfunction2-1}
\end{equation}
In practice, the controls are deterministic. 
Hence, $\textit{E}\big[\mathbf{u}_k^2\big]=\mathbf{u}_k^2$ 
and (\ref{eqn:costfunction2-1}) becomes
\begin{equation}
\begin{aligned}
&\textit{E}\big[\mathcal{J}(\mathbf{x}_k,\mathbf{u}_{k-1})\big]\\
=&\sum_{i=1}^{H}\Big\{\textit{E}\bigg[\big\|\mathbf{x}_{k+i}-\mathbf{r}_{k+i}\big\|^2_{\mathbf{Q}}\bigg]+\big\|\mathbf{u}_{k+i-1}\big\|^2_{\mathbf{R}}\Big\}\\
=&\sum_{i=1}^{H}\Big\{\big\|\meanvalue_{k+i}-\mathbf{r}_{k+i}\big\|^2_{\mathbf{Q}}+\big\|\mathbf{u}_{k+i-1}\big\|^2_{\mathbf{R}}+\textit{trace}\big(\mathbf{Q}\varvalue_{k+i}\big)\Big\} \\
=& h \left(\meanvalue_{k}, \mathbf{u}_{k-1} \right)
\end{aligned}
\label{eqn:costfunction2-2}
\end{equation}

Now we have a deterministic cost function which involve the model variance $\varvalue$ 
that allows model uncertainties to be explicitly included in the computation of the optimized controls.
Note that for multiple-step predictions with uncertainty propagation, the computational complexity of problem (\ref{eqn:optproblem2}) will not increase even though the~\gls{gp} model becomes more complicated. 
This is because the modelling and the control processes are independent of each other.

\subsubsection{Nonlinear Optimization Solution}
With the cost function (\ref{eqn:costfunction2-2}) and the state constraints (\ref{eqn:deterministicConstraints}),
the original stochastic optimization problem (\ref{eqn:optproblem2}) has been relaxed to a deterministic constrained nonlinear optimization problem. But it is typically non-convex.
This is usually solved by derivative-based approaches,
such as Lagrange multipliers \cite{A-Troltzsch-LagranMultipler-2005} based on first-order derivatives (gradient),
or by \gls{sqp} and interior-point algorithms based on second-order derivatives 
(Hessians matrix) \cite{IP-Diehl-EfficientSolution4MPC-2009}.
When the derivative of the cost function is unavailable or is too difficult to compute,
it could be approximated iteratively by a sampling method~\cite{A-Lucidi-DeviativeFreeOpt-2002,A-Liuzzi-SequentialDerivativeFreeOpt-2010}.
Alternatively,
evolutionary algorithms, such as~\gls{pso}~\cite{A-Luo-PSO4NLP-2007} and
\gls{ga}~\cite{A-Yeniay-GA4NLP-2005}, could be used to solve the problem.
This approach is able to handle general constrained optimization problems.
However, there is no guarantee that the solutions obtained are near optimum.
A review of nonlinear optimization techniques for the \gls{mpc} problem can be found in~\cite{IP-Diehl-EfficientSolution4MPC-2009}.

A suitable technique for solving our \gls{mpc} problem is the~\gls{fpsqp} 
algorithm proposed in~\cite{A-Wright-FTRSQP-2004}.
It can be explained using the following general form of a constrained nonlinear optimization problem:
\begin{subequations}\label{eqn:condenseProb}
	\begin{align}
	\min_{\mathbf{z}} \: & h(\mathbf{z}) \qquad \mathbf{z}\in\mathbb{R}^{n+m} \label{subeqn:condenseObj}\\
	\mbox{s.t.} \quad  & c(\mathbf{z}) = 0\\
	& d(\mathbf{z})\leq 0
	\end{align}
\end{subequations}
where $h:\mathbb{R}^{n+m}\rightarrow\mathbb{R}$ is the objective function,
$c:\mathbb{R}^{n+m}\rightarrow\mathbb{R}^{n}$ and $d:\mathbb{R}^{n+m}\rightarrow\mathbb{R}^{n+m}$ represents the corresponding equality and inequality constraints, respectively.
\gls{fpsqp} generates a sequence of feasible solutions $\{\mathbf{z}^j\}_{j=0,1,2,\cdots}$ 
by splitting the original problem into several~\gls{qp} sub-problems.
In particular,
a step $\Delta\mathbf{z}^j$ from current iterate $\mathbf{z}^j$ to the next one $\mathbf{z}^{j+1}$ can be obtained by solving the following \gls{qp} subproblem:
\begin{subequations}\label{eqn:sqpprob}
	\begin{align}
	\min_{\Delta\mathbf{z}^j} &\triangledown h(\mathbf{z}^j)^T\Delta\mathbf{z}^j+\frac{1}{2}{\Delta\mathbf{z}^j}^T\mathbf{H}^j\Delta\mathbf{z}^j\\
	\label{subeqn:C_E}
	\mbox{s.t.}\; &c(\mathbf{z}^j)+\triangledown c(\mathbf{z}^j)^T\Delta\mathbf{z}^j=0\\
	\label{subeqn:C_I}
	& d(\mathbf{z}^j)+\triangledown d(\mathbf{z}^j)^T\Delta\mathbf{z}^j \leq 0
	\end{align}
\end{subequations}
under the trust-region constraint 
\begin{equation}
\big\| \Delta\mathbf{z}^j \big\| \leq \gamma^j 
\end{equation}
where $\triangledown h(\cdot)$ denotes the first-order derivative of the objective function at $\mathbf{z}^j$,
$\triangledown c(\cdot)$ and $\triangledown d(\cdot)$ are two linearised Jacobian matrices at the $\mathbf{z}^j$.
The matrix $\mathbf{H}^j\in\mathbb{R}^{(n+m)\times(n+m)}$ is an exact or approximated Lagrangian Hessian matrix
and $\gamma^j$ represents the trust-region radius.
To guarantee the feasibility of $\Delta\mathbf{z}^j$,
its corresponding perturbation $\Delta\tilde{\mathbf{z}}^j$ which satisfies the following conditions 
need to be computed:
\begin{subequations}
	\begin{align}
	&\mathbf{z}^j+\Delta\tilde{\mathbf{z}}^j\in\bm{\Pi}\\
	&\frac{1}{2}\left\|\Delta\mathbf{z}\right\|_2\leq\left\|\tilde{\Delta\mathbf{z}}\right\|_2\leq\frac{3}{2}\left\|\Delta\mathbf{z}\right\|_2
	\end{align}
\end{subequations}
where $\bm{\Pi}$ denotes the feasible points set of problem (\ref{eqn:condenseProb}).
A method to obtain such a perturbation is proposed in~\cite{A-Peng-FeasibleTRSQP-2006}.
An acceptability value of $\Delta\mathbf{z}^j$ defined by:
\begin{equation}
\rho^j = \frac{h(\mathbf{z}^{j+1})-h(\mathbf{z}^j)}{-\triangledown h(\mathbf{z}^j)^T\Delta\mathbf{z}^j-\frac{1}{2}{\Delta\mathbf{z}^j}^T\mathbf{H}^j\Delta\mathbf{z}^j}
\label{eqn:acceptability}
\end{equation}
If this value is not acceptable,
then the trust-region radius $\gamma^j$ will need to be adjusted.
An adaptive method to adjust $\gamma^j$ can be found in~\cite{A-Zhang-AdaptiveTrustRegion-2002}.
The complete \gls{fpsqp} algorithm is described in Algorithm~\ref{alg:trsqp}.

\IncMargin{1em}
\newcommand{\nosemic}{\renewcommand{\@endalgocfline}{\relax}}% Drop semi-colon ;
\newcommand{\dosemic}{\renewcommand{\@endalgocfline}{\algocf@endline}}% Reinstate semi-colon ;
\newcommand{\pushline}{\Indp}% Indent
\newcommand{\popline}{\Indm\dosemic}% Undent
\let\oldnl\nl% Store \nl in \oldnl
\newcommand{\nonl}{\renewcommand{\nl}{\let\nl\oldnl}}% Remove line number for one line
\renewcommand{\KwIn}{\textbf{Initialization}}
\begin{algorithm}[!t]
	\KwIn{}\\
	\nonl \quad feasible point $\mathbf{z}^0\in\bm{\Pi}$, \\
	\nonl \quad Hessian matrix $\mathbf{H}^0$,\\
	\nonl \quad trust region upper bound $\gamma_{\text{max}}> 0$, \\
	\nonl \quad initial trust region radius $\gamma^0=\| \triangledown h(\mathbf{z}^0)\|$\\
	\nonl \quad $\tau=0$, $0< \tau_1 < \tau_2 < 1$\\ 
          
	\BlankLine
	\For {$j=0,1,2,\cdots, J<\infty$}{
	    Obtain step $\Delta\mathbf{z}^j$ by solving the problem (\ref{eqn:sqpprob});\\
	    \eIf{$\triangledown h(\mathbf{z}^j)^T\Delta\mathbf{z}^j+\frac{1}{2}{\Delta\mathbf{z}^j}^T\mathbf{H}^j\Delta\mathbf{z}^j=0$}{Stop;}{
	    Update $\rho^j$ by using (\ref{eqn:acceptability});\\
	    Update $\mathbf{z}^{j+1}$, $\tau$:
	    \begin{equation}
	    \nonumber
	    \mathbf{z}^{j+1} =\Big\{\begin{array}{ll}
	    \mathbf{z}^{j}+\Delta\mathbf{z}^j, & \rho^j \geq\tau_1 \\
	    \mathbf{z}^{j}, & \text{otherwise}
	    \end{array}
	    \end{equation}
	    \begin{equation}
	    \nonumber
	    \tau =\{\begin{array}{ll}
	    \frac{\|\Delta\mathbf{z}^j\|}{\|\triangledown h(\mathbf{z}^{j+1})-h(\mathbf{z}^j)\|}, & \rho^j \geq\tau_1 \\
	    \tau/4, & \text{otherwise}
	    \end{array}
	    \end{equation}\\
	    Update trust region radius:
	    \begin{equation}
	    \nonumber
	    \gamma^{j+1} =\{\begin{array}{ll}
	    \min\big\{\tau\|\triangledown h(\mathbf{z}^{j+1})\|, \gamma_{\text{max}}\big\}, & \rho^j \geq\tau_2 \\
	    \tau\|\triangledown h(\mathbf{z}^{j+1})\|, & \text{otherwise}
	    \end{array}
	    \end{equation}\\
	    Update Hessian matrix $\mathbf{H}^{j+1}$ by using (\ref{eqn:hessianupdate});\\
	    $j=j+1;$}
    }
\caption{The Feasibility-Perturbed Sequential Quadratic Programming used in the GPMPC1 algorithm}
\label{alg:trsqp}
\end{algorithm}
\DecMargin{1em}

\subsubsection{Application to GPMPC1}
% $\mathbf{z}=[\meanvalue_k^T,\Delta\mathbf{u}_k^T, \sqrt{\text{diag}(\varvalue_k)}^T]^T\in \mathbb{R}^{2n+m}$
Applying \gls{fpsqp} to the GPMPC1 problem (\ref{eqn:optproblem2}),
it should be noted that the constraints (\ref{eqn:deterministicConstraints}) are linear.
Therefore it is possible to simply use $\Delta\tilde{\mathbf{z}}^j=\Delta\mathbf{z}^j$.
The next iterate then can be obtained by
\begin{equation}
\mathbf{z}^{j+1}=\mathbf{z}^j+\Delta\mathbf{z}^j
\label{eqn:nextiterate}
\end{equation}

Expressing the cost function (\ref{eqn:costfunction2-2}) as
(\ref{subeqn:condenseObj}), define 
$\mathbf{z}_k=[\meanvalue_{k}^T,\mathbf{u}_{k-1}^T]^T\in \mathbb{R}^{n+m}$.
Hence,
\begin{equation}\label{eqn:reformulatedObj}
h(\mathbf{z_k})=\sum_{i=1}^{H}\Big\{\mathbf{z}_{k+i}^T
\left[
\begin{array}{cc}
\mathbf{Q} & 0\\
0 & \mathbf{R}
\end{array}
\right] 
\mathbf{z}_{k+i}+\text{trace}\big\{\mathbf{Q}\varvalue_{k+i}\big\}\Big\}
\end{equation}

One key issue in using~\gls{fpsqp} is the local linearisation at $\Delta\mathbf{z}^j$.
The basic~\gls{gp} based local model derived   Section~\ref{subsec:basicLinear} shall be used to derive
the \gls{qp} subproblem as:
\begin{subequations}
	\begin{align}
	\min_{\Delta\mathbf{z}_k,\Delta\varvalue_k}&\sum_{i=1}^{H}
	\Big\{\frac{\partial h}{\partial\mathbf{z}_{k+i}}\Delta\mathbf{z}_{k+i}+\frac{1}{2}\Delta\mathbf{z}_{k+i}^T\mathbf{H}_k\Delta\mathbf{z}_{k+i}\\
	&+\text{trace}\big\{\mathbf{Q}(\varvalue_{k+i}+\Delta\varvalue_{k+i})\big\}
	\Big\}\\
	\mbox{s.t.}\quad  & \Delta\meanvalue_{k+i+1}=\mathbf{A}_{k+i}\Delta\meanvalue_{k+i}+\mathbf{B}_{k+i}\Delta\mathbf{u}_{k+i}\\
	&\mathbf{u}_{\text{min}}\leq\mathbf{u}_{k+i}+\Delta\mathbf{u}_{k+i}\leq\mathbf{u}_{\text{max}}\\
	&\meanvalue_{k+i}+\Delta\meanvalue_{k+i}-2(\varvalue_{k+i}+\Delta\varvalue_{k+i})\geq \mathbf{x}_{\text{min}}\\
	&\meanvalue_{k+i}+\Delta\meanvalue_{k+i}+2(\varvalue_{k+i}+\Delta\varvalue_{k+i}) \leq \mathbf{x}_{\text{max}}\\
	&\left\|\Delta\mathbf{z}_{k+i}\right\| \leq \gamma
	\end{align}
\end{subequations}
Note that
$\mathbf{A}_{k+i}=\frac{\partial\meanvalue_{k+i+1}}{\partial\meanvalue_{k+i}}$ and $\mathbf{B}_{k+i}=\frac{\partial\meanvalue_{k+i+1}}{\partial\mathbf{u}_{k+i}}$ are the two Jacobian matrices of the basic~\gls{gp} based local model (\ref{eqn:gpdynamicmodel-1}).

The computation of the Hessian matrix $\mathbf{H}_k$ of the Lagrangian in (\ref{eqn:sqpprob}) is another key issue when using the \gls{fpsqp} algorithm.
The exact Hessian matrix is usually obtained by
\begin{equation}\label{eqn:exactHessian}
\mathbf{H}_k=\triangledown^2h(\mathbf{z}_k)+\sum_{i=1}^{n}\alpha_i\triangledown^2c(\mathbf{z}_k)+\sum_{i=1}^{n+m}\beta_i\triangledown^2d(\mathbf{z}_k)
\end{equation}
where $\alpha$ and $\beta$ are two Lagrange multipliers applied to the equality and the inequality constraints respectively.
This allows rapid local convergence but requires the second-order derivatives $\triangledown^2c(\mathbf{z}_k)$ which are generally not available.
When the system is represented by a \gls{gp} model,
these derivatives are mathematically computable~\footnote{As shown in~\cite{PHD-Deisenroth-EfficientRLusingGP-2010}, the first-order derivatives are functions of $\tilde{\meanvalue}$ and $\tilde{\varvalue}$, the second-order derivatives therefore can be obtained by using the chain-rule.} but are computationally expensive to obtain.
In addition, 
the exact Hessian matrix may be not positive definite.
To address these issues,
approximation approaches have been proposed in~\cite{A-Tenny-FPSQP4NLP-2004}.
In our work,
$\mathbf{H}_k$ is approximately updated by using a Quasi-Newton method based on the~\gls{bfgs}.
The update equation is given by
\begin{equation}\label{eqn:hessianupdate}
\mathbf{H}_{k+1} = \mathbf{H}_k-\frac{\mathbf{H}_k\Delta\mathbf{z}_k\Delta\mathbf{z}^T_k\mathbf{H}_k}{\Delta\mathbf{z}^T_k\mathbf{H}_k\Delta\mathbf{z}_k}+
\frac{\mathbf{y}_k\mathbf{y}_k^T}{\mathbf{y}_k^T\Delta\mathbf{z}_k}
\end{equation}
where $\Delta\mathbf{z}_k = \mathbf{z}_{k+1}-\mathbf{z}_k$ and $\mathbf{y}_k=\meanvalue_{k+1}-\meanvalue_k$.

\subsection{GPMPC2}
\label{sec:gpmpc2}

With GPMPC1, model uncertainty was introduced through the variance term into the objective function in (\ref{eqn:costfunction2-2}).
But this is an indirect way to handle model uncertainties.
A more direct approach is to introduce the variance into 
that state variable.
This can be done through the use of the extended~\gls{gp} 
based local model (\ref{eqn:gplocalmodel-2}).
In this way, the variances are directly handled in the optimization process.

Another disadvantage of GPMPC1 is that the \gls{mpc} optimization problem (\ref{eqn:optproblem2}) is non-convex.
Due to the recursive nature of \gls{sqp} optimizations,
the process could be time consuming.
With GPMPC2,
the non-convex problem is relaxed to a convex one, 
making it much easier to solve. Sensitivity to initial conditions is reduced and in most cases exact solutions can be obtained \cite{IP-Diehl-EfficientSolution4MPC-2009}.
This convex optimization problem can be solved offline by using \gls{mpqp}~\cite{A-Bemporad-ExplicitLinearQuadRegulorMPC-2002} where the explicit solutions are computed as a lookup table of nonlinear controllers. An example can be found in~\cite{IP-Grancharova-ExplicitMPC-2007}.
However, the size of the table grows exponentially with the number of states. 
Hence it is only suitable for problems with less than 5 states~\cite{A-Wang-FastMPC-2010}.
Using the extended \gls{gp} based local model (\ref{eqn:gplocalmodel-2}), the problem can be solved efficiently by an online active-set algorithm.

\subsubsection{Problem Reformulation}
Based on the extended local model in Section~\ref{subsec:extendedLinear},
define the state variable as
\begin{eqnarray}
	\mathbf{Z}_{k+1} & = & \left[ \mathbf{s}_{k+1|k},\cdots, \mathbf{s}_{k+H|k}\right]^T\in\mathbb{R}^{H(n+n^2)} \nonumber\\
	&=&[\meanvalue_{k+1},\sqrt{\varvalue_{k+1}},\cdots,\meanvalue_{k+H},\sqrt{\varvalue_{k+H}}]^T
\end{eqnarray}
Also, let
\begin{eqnarray}
\mathbf{U}_k&=& \left[ \mathbf{u}_k,\cdots,\mathbf{u}_{k+H-1} \right]^T\in\mathbb{R}^{Hm}\\
\mathbf{r}_{k+1}^*&=& \left[ \mathbf{r}_{k+1},\mathbf{0},\cdots,\mathbf{r}_{k+H},\mathbf{0}\right]^T\in\mathbb{R}^{H(n+n^2)}
\end{eqnarray}
Problem (\ref{eqn:optproblem2}) then becomes
\begin{subequations}\label{eqn:optproblem3-1}
\begin{align}
\min_{\mathbf{U}}&\left\{ \left\|\mathbf{Z}_{k+1}-\mathbf{r}_{k+1}^*\right\|^2_{\tilde{\mathbf{Q}}}+\left\|\mathbf{U}_{k+1}\right\|^2_{\tilde{\mathbf{R}}} \right\}\\
\mbox{s.t.}\quad
&\mathbf{I}_{Hn}\mathbf{x}_{\text{min}}\leq\mathbf{M}_{z}\mathbf{Z}_{k+1}\leq\mathbf{I}_{Hn}\mathbf{x}_{\text{max}}\\
&\mathbf{I}_{Hm}\mathbf{u}_{\text{min}}\leq\mathbf{U}_{k+1}\leq\mathbf{I}_{Hm}\mathbf{u}_{\text{max}}
\end{align}
\end{subequations}
where
\begin{subequations}
\begin{align}
\tilde{\mathbf{Q}}= &\text{diag}\{[\mathbf{Q},\text{diag}\{\textbf{vec}(\mathbf{Q})\},\cdots,\mathbf{Q},\text{diag}\{\textbf{vec}(\mathbf{Q})\}]\}\in\mathbb{R}^{H(n+n^2)\times H(n+n^2)}, \\
\tilde{\mathbf{R}}=&\text{diag}\{[\mathbf{R},\cdots,\mathbf{R}]\} \in\mathbb{R}^{Hm\times Hm},
\end{align}
\end{subequations}
$\mathbf{I}_{a}\in\mathbb{R}^{a}$ is the identity vector,
and
\begin{equation}
\mathbf{M}_z=\left[ 
\begin{array}{cccccc}
\mathbf{I}^T_{n}& 2\mathbf{I}^T_{n^2}& \mathbf{0}&\mathbf{0}&\cdots &\mathbf{0}\\
\mathbf{0}& \mathbf{0}& \mathbf{I}^T_{n}& 2\mathbf{I}^T_{n^2} &\cdots &\mathbf{0}\\
\vdots& \vdots& \vdots& \vdots& \vdots& \vdots\\
\mathbf{0}& \mathbf{0}& \mathbf{0} & \cdots &\mathbf{I}^T_{n}& 2\mathbf{I}^T_{n^2}
\end{array}\right]\in\mathbb{R}^{H\times H(n+n^2)}
\end{equation}
Let $\mathbf{T}_u\in\mathbb{R}^{Hm\times Hm}$ be a lower triangular matrices with unit entries.
Then,
\begin{equation}\label{eqn:linearU} 
\mathbf{U}_k=\mathbf{I}_{Hm}\mathbf{u}_{k-1}+\mathbf{T}_u\Delta\mathbf{U}_k
\end{equation}
$\Delta\mathbf{Z}_{k+1}$ can be expressed as
\begin{equation}
\Delta\mathbf{Z}_{k+1}=\tilde{\mathbf{A}}\Delta\mathbf{s}_{k}+\tilde{\mathbf{B}}\Delta\mathbf{U}_k
\end{equation}
based on the extended local model, 
with the state and control matrices given by
\begin{subequations}
\begin{align}
\tilde{\mathbf{A}}=&\left[\mathbf{A}, \mathbf{A}^2, \cdots, \mathbf{A}^H\right]^T\in\mathbb{R}^{H(n+n^2)}\\
\tilde{\mathbf{B}}=&\left[\begin{array}{cccc}
\mathbf{B} &\mathbf{0}  &\cdots &\mathbf{0}\\
\mathbf{AB}&\mathbf{B}  &\cdots &\mathbf{0}\\
\vdots     & \vdots & \vdots & \vdots\\
\mathbf{A}^{H-1}\mathbf{B}&\mathbf{A}^{H-2}\mathbf{B}&\cdots&\mathbf{B}
\end{array}\right] \in\mathbb{R}^{H(n+n^2)\times Hm}
\end{align}
\end{subequations}
where $\mathbf{A}$ and $\mathbf{B}$ are the two Jacobian matrices 
(\ref{eqn:extendgpJacobian}) and (\ref{eqn:computationextendgpJacobian}) respectively.
The corresponding state variable $\mathbf{Z}_{k+1}$ is therefore given by
\begin{equation}\label{eqn:linearZ}
\mathbf{Z}_{k+1}=\mathbf{s}_k+\mathbf{T}_z\left(\tilde{\mathbf{A}}\Delta\mathbf{s}_{k}+\tilde{\mathbf{B}}\Delta\mathbf{U}_k\right)
\end{equation}
where $\mathbf{T}_z\in\mathbb{R}^{H(n+n^2)\times H(n+n^2)}$ denotes a lower triangular matrix with unit entries.

Based on (\ref{eqn:linearU}) and (\ref{eqn:linearZ}),
problem (\ref{eqn:optproblem3-1}) can be expressed in a more compact form as
\begin{subequations}\label{eqn:optproblem3-2}
\begin{align}
\min_{\Delta\mathbf{U}}&\frac{1}{2}\left\| \Delta\mathbf{U}_k\right\|^2_{\bm{\Phi}}+\bm{\psi}^T\Delta\mathbf{U}_k+\mathbf{C}\\
\mbox{s.t.} \quad 
&\label{eqn:optproblem3-2-constraints}\Delta\mathbf{U}_{\text{min}}\leq
\left[\begin{array}{c}
\mathbf{T}_u \\\mathbf{T}_z\tilde{\mathbf{B}}
\end{array}\right]
\Delta\mathbf{U}_k\leq\Delta\mathbf{U}_{\text{max}}
\end{align}
\end{subequations}
where
\begin{subequations}\label{eqn:condenseConstraints}
\begin{align}
\bm{\Phi}=&\tilde{\mathbf{B}}^T\mathbf{T}_z^T\tilde{\mathbf{Q}}\mathbf{T}_z\tilde{\mathbf{B}}+\mathbf{T}_u^T\tilde{\mathbf{R}}\mathbf{T}_u\in\mathbb{R}^{Hm\times Hm}\\
\bm{\psi}=&2(\mathbf{s}_k\tilde{\mathbf{Q}}\mathbf{T}_z\tilde{\mathbf{B}}+\Delta\mathbf{s}_k\tilde{\mathbf{A}}^T\tilde{\mathbf{Q}}\tilde{\mathbf{B}}\\
&-\mathbf{r}_{k+1}^*\tilde{\mathbf{Q}}\mathbf{T}_z\tilde{\mathbf{B}}+\mathbf{u}_{k-1}\tilde{\mathbf{R}}\mathbf{T}_u)\in\mathbb{R}^{Hm}\\
\mathbf{C}=&(\mathbf{s}_k^2+\mathbf{r}^*_{k+1})\tilde{\mathbf{Q}}+2\mathbf{s}_k\Delta\mathbf{s}_k\tilde{\mathbf{Q}}\mathbf{T}_z\tilde{\mathbf{A}}\\
&+\mathbf{u}_{k-1}^2\tilde{\mathbf{R}}+\Delta\mathbf{s}_k^2\tilde{\mathbf{A}}^T\tilde{\mathbf{Q}}\tilde{\mathbf{A}}\\
&-2\mathbf{r}_{k+1}^*(\mathbf{s}_k\tilde{\mathbf{Q}}-\Delta\mathbf{s}_k\tilde{\mathbf{Q}}\mathbf{T}_z\tilde{\mathbf{A}})\\
\Delta\mathbf{U}_{\text{min}}
=&\left[\begin{array}{c}
\mathbf{I}_{Hm}(\mathbf{u}_{\text{min}}-\mathbf{u}_{k-1})\\
\mathbf{I}_{H(n+n^2)}(\mathbf{x}_{\text{min}}-\mathbf{s}_k-\mathbf{T}_z\tilde{\mathbf{A}}\Delta\mathbf{s}_k)
\end{array}\right]\\
\Delta\mathbf{U}_{\text{max}}
=&\left[\begin{array}{c}
\mathbf{I}_{Hm}(\mathbf{u}_{\text{max}}-\mathbf{u}_{k-1})\\
\mathbf{I}_{H(n+n^2)}(\mathbf{x}_{\text{max}}-\mathbf{s}_k-\mathbf{T}_z\tilde{\mathbf{A}}\Delta\mathbf{s}_k)
\end{array}\right]
\end{align}
\end{subequations}
Since $\tilde{\mathbf{Q}}, \tilde{\mathbf{R}}, \mathbf{T}_z$ and $\mathbf{T}_u$ are positive definite,
$\bm{\Phi}$ is also positive definite.
Hence (\ref{eqn:optproblem3-2}) is a constrained~\gls{qp} problem and is strictly convex.
The solution will therefore be unique and satisfies the~\gls{kkt} conditions.

\subsubsection{Quadratic Programming Solution}
\label{subsec:activeset}
The optimization problem (\ref{eqn:optproblem3-2}) can be solved by 
an active-set method~\cite{B-Fletcher-PracticalMethods4Opt-1987}.
It iteratively seeks an active (or working) set of constraints
and solve an equality constrained~\gls{qp} problem until the optimal solution is found.
The advantage of this method is that accurate solutions can still be obtained even when they are ill-conditioned or degenerated.
In addition, it is conceptually simple and easy to implement.
A warm-start technique could also be used
to accelerate the optimization process substantially.

Let $\mathbf{G}=[\mathbf{T}_u,\mathbf{T}_z\tilde{\mathbf{B}}]^T$,
the constraint (\ref{eqn:optproblem3-2-constraints}) 
can be written as
\begin{equation}\label{eqn:inequalityConstraints}
\left[\begin{array}{c}
\mathbf{G}\\
-\mathbf{G}
\end{array}\right]\Delta\mathbf{U}
\leq\left[ \begin{array}{c}
\Delta\mathbf{U}_{\text{max}}\\
-\Delta\mathbf{U}_{\text{min}}
\end{array}
\right]
\end{equation}
Ignoring the constant term $\mathbf{C}$, problem (\ref{eqn:optproblem3-2}) becomes
\begin{subequations}\label{eqn:optproblem3-3}
\begin{align}
\min_{\Delta\mathbf{U}}&\frac{1}{2}\left\| \Delta\mathbf{U}_k\right\|^2_{\bm{\Phi}}+\bm{\psi}^T\Delta\mathbf{U}_k\\
&\mbox{s.t.} \quad
\label{eqn:optproblem3-3-constraints}\tilde{\mathbf{G}}\Delta\mathbf{U}_k\leq\tilde{\Delta}_\mathbf{U}
\end{align}
\end{subequations}
where $\tilde{\mathbf{G}}=[\mathbf{G}, -\mathbf{G}]^T\in\mathbb{R}^{2H(m+n+n^2)\times Hm}$ and $\tilde{\Delta}_\mathbf{U}=[\Delta\mathbf{U}_{\text{max}}, -\Delta\mathbf{U}_{\text{min}}]^T\in\mathbb{R}^{2H(m+n+n^2)}$.

Let $\bm{\Pi}_{\Delta\mathbf{U}}$ be the set of feasible points,
and $\mathcal{I}=\{1,\cdots,2H(m+n+n^2)\}$ be the constraint index set.
For a feasible point $\Delta\mathbf{U}_k^*\in\bm{\Pi}_{\Delta\mathbf{U}}$,
the index set for the active set of constraints is defined as
\begin{equation}\label{eqn:activeset}
\mathcal{A}(\Delta\mathbf{U}_k^*)=\{i\subseteq\mathcal{I}|\tilde{\mathbf{G}}_{i}\Delta\mathbf{U}_k^*=\tilde{\Delta}_{\mathbf{U},i}\}
\end{equation}
where $\tilde{\mathbf{G}}_i$ is the $i^{th}$ row of  $\tilde{\mathbf{G}}$
and $\tilde{\Delta}_{\mathbf{U},i}$ is the $i^{th}$ row of the $\tilde{\Delta}_{\mathbf{U}}$.
The inactive set is therefore given by
\begin{equation}\label{eqn:inactiveset}
\begin{aligned}
\mathcal{B}(\Delta\mathbf{U}_k^*)
&=\mathcal{I}\setminus\mathcal{A}(\Delta\mathbf{U}_k^*)\\
&=\{i\subseteq\mathcal{I}|\tilde{\mathbf{G}}_i\Delta\mathbf{U}_k^*<\tilde{\Delta}_{\mathbf{U},i}\}
\end{aligned}
\end{equation}
Given any iteration $j$, 
the working set $\mathcal{W}_k^j$ contains all the equality constraints plus the inequality constraints in the active set.
The following~\gls{qp} problem subject to the equality constraints w.r.t. $\mathcal{W}_k^j$ is considered given the feasible points $\Delta\mathbf{U}_k^j\in\bm{\Pi}_{\Delta\mathbf{U}}$:
\begin{subequations}\label{eqn:optproblem3-4}
\begin{align}
&\min_{\bm{\delta}^j}\frac{1}{2}\left\| \Delta\mathbf{U}_k^j+\bm{\delta}^j\right\|^2_{\bm{\Phi}}+\bm{\psi}^T(\Delta\mathbf{U}_k^j+\bm{\delta}^j)\\
=&\min_{\bm{\delta}^j}\frac{1}{2}\left\|\bm{\delta}^j\right\|^2_{\bm{\Phi}}+(\bm{\psi}+\bm{\Phi}\Delta\mathbf{U}^j_k)^T\bm{\delta}^j\\
\nonumber&+\underbrace{\frac{1}{2}\left\| \Delta\mathbf{U}_k^j\right\|^2_{\bm{\Phi}}+\bm{\psi}^T\Delta\mathbf{U}_k^j}_{\text{constant}}\\
&\mbox{s.t.} \quad
\label{eqn:optproblem3-4-constraints}\tilde{\mathbf{G}}_i(\Delta\mathbf{U}_k^j+\bm{\delta}^j)=\tilde{\Delta}_{\mathbf{U},i}, i\in\mathcal{W}_k^j
\end{align}
\end{subequations}
This problem can be simplified by ignoring the constant term to:
\begin{subequations}\label{eqn:optproblem3-5}
\begin{align}
&\min_{\bm{\delta}^j}\frac{1}{2}\left\|\bm{\delta}^j\right\|^2_{\bm{\Phi}}+(\bm{\psi}+\bm{\Phi}\Delta\mathbf{U}^j_k)^T\bm{\delta}^j\\
=&\min_{\bm{\delta}^j}\frac{1}{2}{\bm{\delta}^j}^T\bm{\Phi}\bm{\delta}^j+(\bm{\psi}+\bm{\Phi}\Delta\mathbf{U}^j_k)^T\bm{\delta}^j\\
&\mbox{s.t.} \quad
\label{eqn:optproblem3-5-constraints}\tilde{\mathbf{G}}_i\bm{\delta}^j=\tilde{\Delta}_{\mathbf{U},i}-\tilde{\mathbf{G}}_i\Delta\mathbf{U}_k^j, i\in\mathcal{W}_k^j
\end{align}
\end{subequations}

By applying the~\gls{kkt} conditions to problem (\ref{eqn:optproblem3-5}),
we can obtain the following linear equations: 
\begin{equation}\label{eqn:activeset-kkt-2}
\underbrace{\left[\begin{array}{cc}
\bm{\Phi} & \tilde{\mathbf{G}}_{\mathcal{A}}^T\\
\tilde{\mathbf{G}}_{\mathcal{A}} & \mathbf{0}
\end{array}\right]}_{\text{Lagrangian Matrix}}
\left[\begin{array}{c}
\bm{\delta}^j \\ \bm{\lambda}^*_k
\end{array}\right]=
\left[\begin{array}{c}
-\bm{\psi}-\bm{\Phi}\Delta\mathbf{U}^j_k\\ 
\tilde{\Delta}_{\mathbf{U},\mathcal{A}}-\tilde{\mathbf{G}}_{\mathcal{A}}\Delta\mathbf{U}_k^j
\end{array}\right]
\end{equation}
where $\bm{\lambda}_k^*\in\mathbb{R}^{2H(m+n+n^2)}$ denotes the vector of Lagrangian multipliers,
$\tilde{\mathbf{G}}_{\mathcal{A}}\subseteq\tilde{\mathbf{G}}$ and $\tilde{\Delta}_{\mathbf{U},\mathcal{A}}\subset\tilde{\Delta}_\mathbf{U}$ are the weighting matrix and the upper bounds of the constraints w.r.t.  $\mathcal{W}_k^j$.
Let the inverse of Lagrangian matrix be denoted by
\begin{equation}
\left[\begin{array}{cc}
\bm{\Phi} & \tilde{\mathbf{G}}_{\mathcal{A}}^T\\
\tilde{\mathbf{G}}_{\mathcal{A}} & \mathbf{0}
\end{array}\right]^{-1}=
\left[\begin{array}{cc}
\mathbf{L}_1 & \mathbf{L}_2^T\\
\mathbf{L}_2 & \mathbf{L}_3\\
\end{array}\right]
\end{equation}
If this inverse exists, then the solution is given by
\begin{subequations}
\begin{align}
\bm{\delta}^j &=-\mathbf{L}_1(\bm{\psi}+\bm{\Phi}\Delta\mathbf{U}^j_k)
+\mathbf{L}_2^T(\tilde{\Delta}_{\mathbf{U},\mathcal{A}}-\tilde{\mathbf{G}}_{\mathcal{A}}\Delta\mathbf{U}_k^j)\\
\bm{\lambda}_k^*&=-\mathbf{L}_2(\bm{\psi}+\bm{\Phi}\Delta\mathbf{U}^j_k)
+\mathbf{L}_3(\tilde{\Delta}_{\mathbf{U},\mathcal{A}}-\tilde{\mathbf{G}}_{\mathcal{A}}\Delta\mathbf{U}_k^j)
\end{align}
\end{subequations}
where
\begin{subequations}
\begin{align}
\mathbf{L}_1&=\bm{\Phi}^{-1}-\bm{\Phi}^{-1}\tilde{\mathbf{G}}^T_\mathcal{A}(\tilde{\mathbf{G}}_\mathcal{A}\bm{\Phi}^{-1}\tilde{\mathbf{G}}^T_\mathcal{A})^{-1}\tilde{\mathbf{G}}_\mathcal{A}\bm{\Phi}^{-1}\\
\mathbf{L}_2&=\bm{\Phi}^{-1}\tilde{\mathbf{G}}^T_\mathcal{A}(\tilde{\mathbf{G}}_\mathcal{A}\bm{\Phi}^{-1}\\
\mathbf{L}_3&=-(\tilde{\mathbf{G}}_\mathcal{A}\bm{\Phi}^{-1}\tilde{\mathbf{G}}^T_\mathcal{A})^{-1}
\end{align}
\end{subequations}

\newcommand{\newargmin}{\mathop{\mathrm{argmin}}}       
If $\bm{\delta}^j\neq 0$,
then the set of feasible points $\Delta\mathbf{U}_k^j$ fails to minimize problem (\ref{eqn:optproblem3-3}).
In this case, the next set of feasible point is computed for the
next iteration by $\Delta\mathbf{U}_k^{j+1}=\Delta\mathbf{U}_k^j+\kappa^j\bm{\delta}^j$
with step size
\begin{equation}\label{eqn:activeset-step}
\kappa^j=\min\left\lbrace 
1, \min_{i\in\mathcal{B}(\Delta\mathbf{U}_k^j)}\frac{\tilde{\Delta}_{\mathbf{U},i}-\tilde{\mathbf{G}}_i\Delta\mathbf{U}_k^j}{\tilde{\mathbf{G}}_i\bm{\delta}^j}
\right\rbrace 
\end{equation}
If $\kappa^j<1$,
the inequality constraint with index $q=\newargmin_{i\in\mathcal{B}(\Delta\mathbf{U}_k^j)}\frac{\tilde{\Delta}_{\mathbf{U},i}-\tilde{\mathbf{G}}_i\Delta\mathbf{U}_k^j}{\tilde{\mathbf{G}}_i\bm{\delta}^j}$ should be ``activated", giving the working set $\mathcal{W}_k^{j+1}=\mathcal{W}_k^j\cup q$.
Otherwise, we have $\mathcal{W}_k^{j+1}=\mathcal{W}_k^j$.

Alternatively, if the solution gives $\bm{\delta}^j=0$,
then the current feasible points $\Delta\mathbf{U}_k^j$ could be the optimal solution.
This can be verified by checking the Lagrangian multiplier $\lambda_k^*=\min_{i\in\mathcal{W}_k^j\cap\mathcal{I}}\bm{\lambda}_{k,i}^*$.
If $\lambda_k^*\geq0$,
the optimal solution of the (\ref{eqn:optproblem3-3}) at sampling time $k$ is found.
Otherwise, this inequality constraint indexed by $p=\newargmin_{i\in\mathcal{W}_k^j\cap\mathcal{I}}\bm{\lambda}_{k,i}^*$ should be removed from the current working set, giving us $\mathcal{W}_k^{j+1}=\mathcal{W}_k^j\setminus p$.
Algorithm~\ref{alg:activeset} summarizes the active set algorithm used in the GPMPC2.

\IncMargin{1em}
\begin{algorithm}[!t]
	\KwIn{}\\
	\nonl \quad the feasible point $\Delta\mathbf{U}_k^0\in\bm{\Pi}_{\Delta\mathbf{U}}$; \\
	\nonl \quad the working set $\mathcal{W}^0=\mathcal{A}(\Delta\mathbf{U}^0_k)$;\\ 
	\BlankLine
	\For {$j=0,1,2,\cdots$}{
	    Compute the $\bm{\delta}^j$ and $\bm{\lambda}_k^*$ by solving the linear equations (\ref{eqn:activeset-kkt-2});\\
	    \eIf{$\bm{\delta}^j=0$}
	    {$\lambda_k^*=\min_{i\in\mathcal{W}_k^j\cap\mathcal{I}}\bm{\lambda}_{k,i}^*$,\\
	     $p=\newargmin_{i\in\mathcal{W}_k^j\cap\mathcal{I}}\bm{\lambda}_{k,i}^*$\\
	    \eIf{$\lambda_k^*\geq 0$}{$\Delta\mathbf{U}_k^*=\Delta\mathbf{U}^j_k$;\\
	       Stop.}
	       {$\mathcal{W}_k^{j+1}=\mathcal{W}_k^j\setminus p$;\\
	       $\Delta\mathbf{U}^{j+1}_k=\Delta\mathbf{U}^j_k$;}
	    }
	    {Compute the step length $\kappa^j$ by (\ref{eqn:activeset-step}),\\
	    $q=\newargmin_{i\in\mathcal{B}(\Delta\mathbf{U}_k^j)}\frac{\tilde{\Delta}_{\mathbf{U},i}-\tilde{\mathbf{G}}_i\Delta\mathbf{U}_k^j}{\tilde{\mathbf{G}}_i\bm{\delta}^j}$\\
	    \eIf{$\kappa^j<1$}
	    {$\Delta\mathbf{U}^{j+1}_k=\Delta\mathbf{U}^j_k+\kappa^j\bm{\delta}^j$;\\ $\mathcal{A}(\Delta\mathbf{U}^{j+1}_k)=\mathcal{A}(\Delta\mathbf{U}^j_k)\cup q$;}
	    {$\Delta\mathbf{U}^{j+1}_k=\Delta\mathbf{U}^j_k+\bm{\delta}^j$;\\ $\mathcal{A}(\Delta\mathbf{U}^{j+1}_k)=\mathcal{A}(\Delta\mathbf{U}^j_k)$;}}
    }
\caption{Active set method for solving the GPMPC2 problem}
\label{alg:activeset}
\end{algorithm}
\DecMargin{1em}

\subsubsection{Implementation Issues}
The key to solving the linear equations (\ref{eqn:activeset-kkt-2}) is the inverse of the Lagrangian matrix.
However, $\tilde{\mathbf{G}}_\mathcal{A}$ is not always full ranked.
Thus the Lagrangian matrix is not always invertible.
This problem can be solved by decomposing $\tilde{\mathbf{G}}_\mathcal{A}$ using QR factorization technique,
giving us
$\mathbf{G}_{\mathcal{A}}^T=\mathcal{Q}
\left[ \mathcal{R} \;\; \mathbf{0} \right]^{T}$
where $\mathcal{R}\in\mathbb{R}^{m_1\times m_1}$ is an upper triangular matrix with $m_1=\text{rank}(\tilde{\mathbf{G}}_\mathcal{A})$.
$\mathcal{Q}\in\mathbb{R}^{Hm\times Hm}$ is an orthogonal matrix that can be further decomposed to $\mathcal{Q}=\left[\mathcal{Q}_1\;\mathcal{Q}_2\right]$ where $\mathcal{Q}_1\in\mathbb{R}^{Hm\times m_1}$ and $\mathcal{Q}_2\in\mathbb{R}^{Hm\times (Hm-m_1)}$.
Thus, $\mathbf{G}_{\mathcal{A}}^T=\mathcal{Q} =\mathcal{Q}_1\mathcal{R}$ and 
\begin{subequations}
\begin{align}
\mathbf{L}_1&=\mathcal{Q}_2(\mathcal{Q}_2^T\bm{\Phi}\mathcal{Q}_2)^{-1}\mathcal{Q}_2^T\\
\mathbf{L}_2&=\mathcal{Q}_1{\mathcal{R}^{-1}}^T-\mathbf{L}_1\bm{\Phi}\mathcal{Q}_1{\mathcal{R}^{-1}}^T\\
\mathbf{L}_3&=\mathcal{R}^{-1}\mathcal{Q}_1^T\bm{\Phi}\mathbf{L}_2
\end{align}
\end{subequations}

The second issue relates to using the appropriate warm-start technique to improve the convergence rate of the active-set method.
For GPMPC2,
since the changes in the state between two successive sampling instants are usually quite small,
we simply use the previous $\Delta\mathbf{U}_k^*$ as the starting point $\Delta\mathbf{U}^0_{k+1}$ for the next sampling time $k+1$.
This warm-start technique is usually employed in \gls{mpc} optimizations because of its proven effectiveness~\cite{A-Wang-FastMPC-2010}.

\subsection{Stability}
\label{subsec:stability}

The stability of the closed-loop controller is not guaranteed because the \gls{mpc} problem is open-loop.
This can be demonstrated by the stability analysis of the proposed algorithms.

In particular,
for the~\gls{mpc} problem (\ref{eqn:optproblem2}) in the GPMPC1 algorithm,
the objective (\ref{eqn:costfunction2-2}) can be directly used as the Lyapunov function.
Therefore, it can be known that
\begin{equation}\label{eqn:lyapunov-k}
\mathcal{V}^*(k)=\sum_{i=1}^{H}\left\lbrace 
\left\|\Delta\meanvalue^*_{k+i}\right\|_\mathbf{Q}^2+ \left\|\mathbf{u}^*_{k+i-1}\right\|_\mathbf{R}^2+\text{trace}(\mathbf{Q}\varvalue^*_{k+i})
\right\rbrace 
\end{equation}
where $\Delta\meanvalue^*_{k+i}=\meanvalue^*_{k+i}-\mathbf{r}_{k+i}$,
$\mathbf{u}^*$ is the optimal control inputs,
and $\meanvalue^*_{k+i}$ and $\varvalue^*_{k+i}$ represent the corresponding optimal means and variances of the~\gls{gp} model at time $k$.
The Lyapunov function at time $k+1$ is subsequently obtained by,
\begin{subequations}\label{eqn:lyapunov-k+1}
\begin{align}
&\mathcal{V}(k+1)\\
&=\sum_{i=1}^{H}\left\lbrace
\left\|\Delta\meanvalue_{k+1+i}\right\|_\mathbf{Q}^2+ \left\|\mathbf{u}_{k+i}\right\|_\mathbf{R}^2+\text{trace}(\mathbf{Q}\varvalue_{k+1+i})
\right\rbrace\\
&=\mathcal{V}^*(k)-\left\|\Delta\meanvalue^*_{k+1}\right\|_\mathbf{Q}^2-\left\|\mathbf{u}^*_k\right\|_\mathbf{R}^2-\text{trace}(\mathbf{Q}\varvalue_{k+1}^*)\\
\nonumber&+\left\|\Delta\meanvalue_{k+1+H}\right\|_\mathbf{Q}^2+\left\|\mathbf{u}_{k+H}\right\|_\mathbf{R}^2+\text{trace}(\mathbf{Q}\varvalue_{k+1+H})
\end{align}
\end{subequations}
It is easy to know that $\mathcal{V}^*(k+1)\leq\mathcal{V}(k+1)$ due to the nature of the optimization.
Furthermore,
the following inequality can be obtained,
\begin{subequations}\label{eqn:lyapunov-minus}
\begin{align}
\mathcal{V}^*(k+1)\leq&\mathcal{V}(k+1)\\
\leq&\mathcal{V}^*(k)+\left\|\Delta\meanvalue_{k+1+H}\right\|_\mathbf{Q}^2\\
\nonumber&+\left\|\mathbf{u}_{k+H}\right\|_\mathbf{R}^2+\text{trace}(\mathbf{Q}\varvalue_{k+1+H})
\end{align}
\end{subequations}
because of $\left\|\Delta\meanvalue^*_{k+1}\right\|_\mathbf{Q}^2\geq 0$,
 $\left\|\mathbf{u}_k^*\right\|_\mathbf{R}^2\geq 0$ and $\text{trace}(\mathbf{Q}\varvalue^*_{k+1})\geq 0$.
The stability result of the problem (\ref{eqn:optproblem3-2}) in the GPMPC2 algorithm can be obtained in the same way.

The result in (\ref{eqn:lyapunov-minus}) shows that,
to guarantee the stability,
additional terminal constraints on the means and variances of the~\gls{gp} model, as well as the control inputs are required such that,
\begin{subequations}\label{eqn:terminalconstraints}
\begin{align}
\meanvalue_{k+H+1|k}-\mathbf{r}_{k+H+1}&=0\\
\varvalue_{k+H+1|k}&=0\\
\mathbf{u}_{k+H|k}&=0
\end{align}
\end{subequations}
However,
it should be noted that,
these newly added constraints altered the optimization problem.
Hence its feasibility will need to be analysed.
Another approach to provide the guaranteed stability is by introducing a terminal cost into the objective function ~\cite{A-Mayne-ConstrainedMPC-StabilityOptimality-2000}.

\section{Numerical Simulations}
\label{sec:simulation}

GPMPC1 and GPMPC2 are applied to two trajectory tracking problems of a~\gls{mimo} nonlinear system with time-varying parameters.
For each problem, $50$ independent simulations are performed on a computer with a $3.40$GHz Intel$\circledR$ Core$^{\text{TM}}$ $2$ Duo CPU with $16$ GB RAM, using Matlab$\circledR$ version $8.1$.
The average simulation results of these $50$ trials are presented here.

%\begin{comment}
%The modelling performance is evaluated by the training~\gls{mse}, defined as
%\begin{equation}
%\nonumber\text{MSE}=\frac{1}{N}\sum_{k=1}^{N}(\mathbf{y}_k-\mathbf{r}_k)^2
%\end{equation}
%where $N$ the total number of time steps.
%The control performance is measured in terms of the computed inputs and outputs,
%as well as the~\gls{iae} value that describes the set-point tracking error.
%\begin{equation}
%\nonumber\text{IAE}=\sum_{k=1}^{N} \left\| \mathbf{y}_k-\mathbf{r}_k \right\|
%\end{equation}
%\end{comment}

The~\gls{mimo} nonlinear system in~\cite{A-Pan-MPCusingANN-2012} is used
for our simulations.
It is described by:
\begin{equation}\label{eqn:nltvsysexample}
\begin{aligned}
x_{1}(k+1) &= \frac{x_{1}(k)^2}{1+x_{1}(k)^2} + 0.3x_{2}(k)\\
x_{2}(k+1) &= \frac{x_{1}(k)^2}{1+x_{2}(k)^2+x_{3}(k)^2+x_{4}(k)^2} \\
     &+ a(k)u_1(k)\\
x_{3}(k+1) &= \frac{x_{3}(k)^2}{1+x_{3}(k)^2} + 0.2x_{4}(k)\\
x_{4}(k+1) &= \frac{x_{3}(k)^2}{1+x_{1}(k)^2+x_{2}(k)^2+x_{4}(k)^2} \\
     &+ b(k)u_2(k)\\
y_1(k+1) &= x_{1}(k+1) + \omega_1\\
y_2(k+1) &= x_{3}(k+1) + \omega_2
\end{aligned}
\end{equation}
where $x_1,x_2,x_3$ and $x_4$ are system states, $u_1,u_2$ and $y_1,y_2$ denote system inputs and outputs, respectively.
$\omega_1,\omega_2\sim \mathcal{N}(0,0.01)$ are independent Gaussian white noise.
In addition, the time-varying parameters $a(k)$ and $b(k)$ are given by
\begin{subequations}
\begin{align}
a(k)&=10+0.5\sin(k)\\
b(k)&=\frac{10}{1+\exp(-0.05k)}
\end{align}
\end{subequations}

\begin{figure}[!t]
\subfigure[``Step" trajectory]{
\centering
\label{fig:trainerror_step}
\includegraphics[width=\singlefigWidth\textwidth]{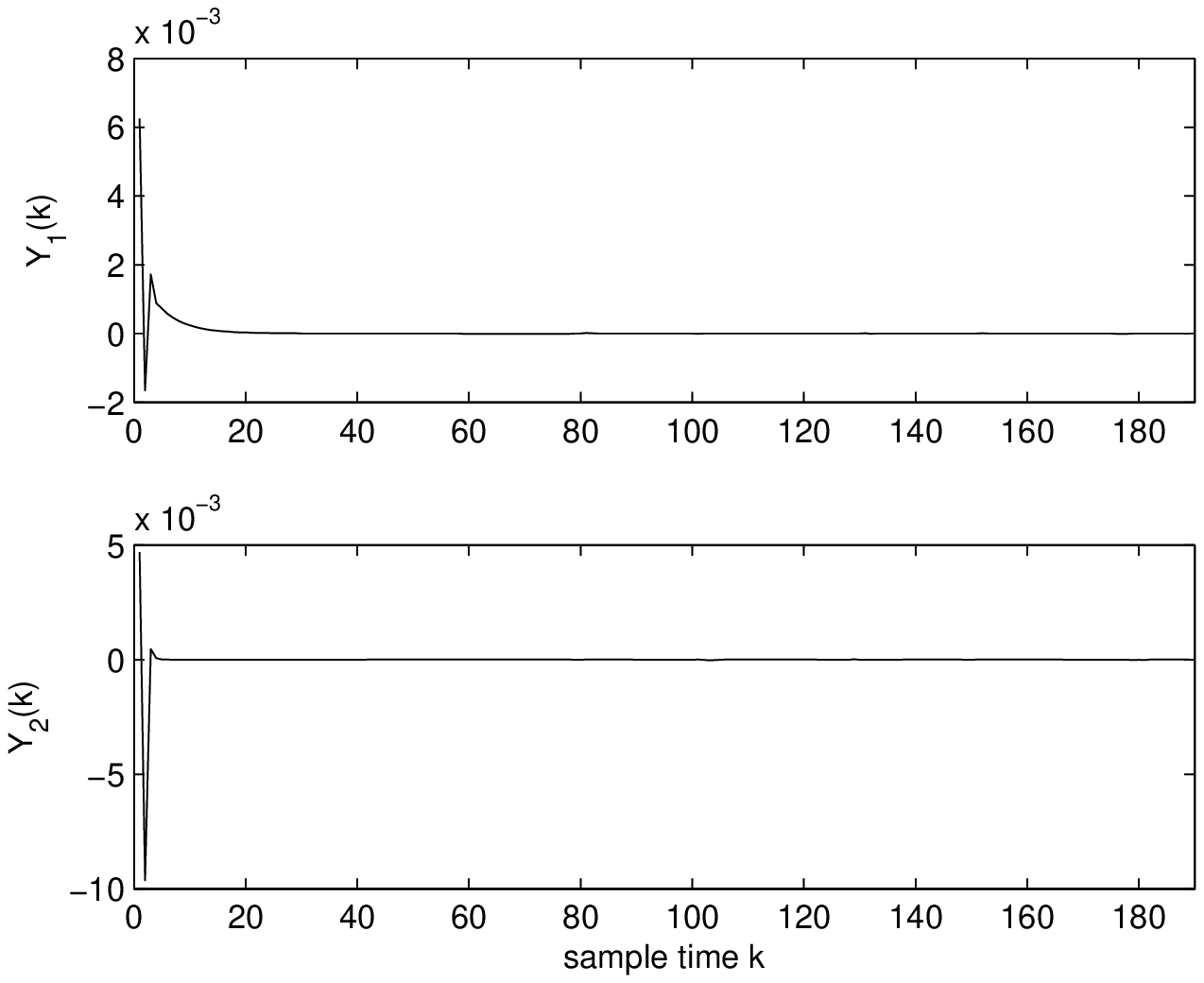}}
\subfigure[``Lorenz" trajectory]{
\label{fig:trainerror_lorenz}
\includegraphics[width=\singlefigWidth\textwidth]{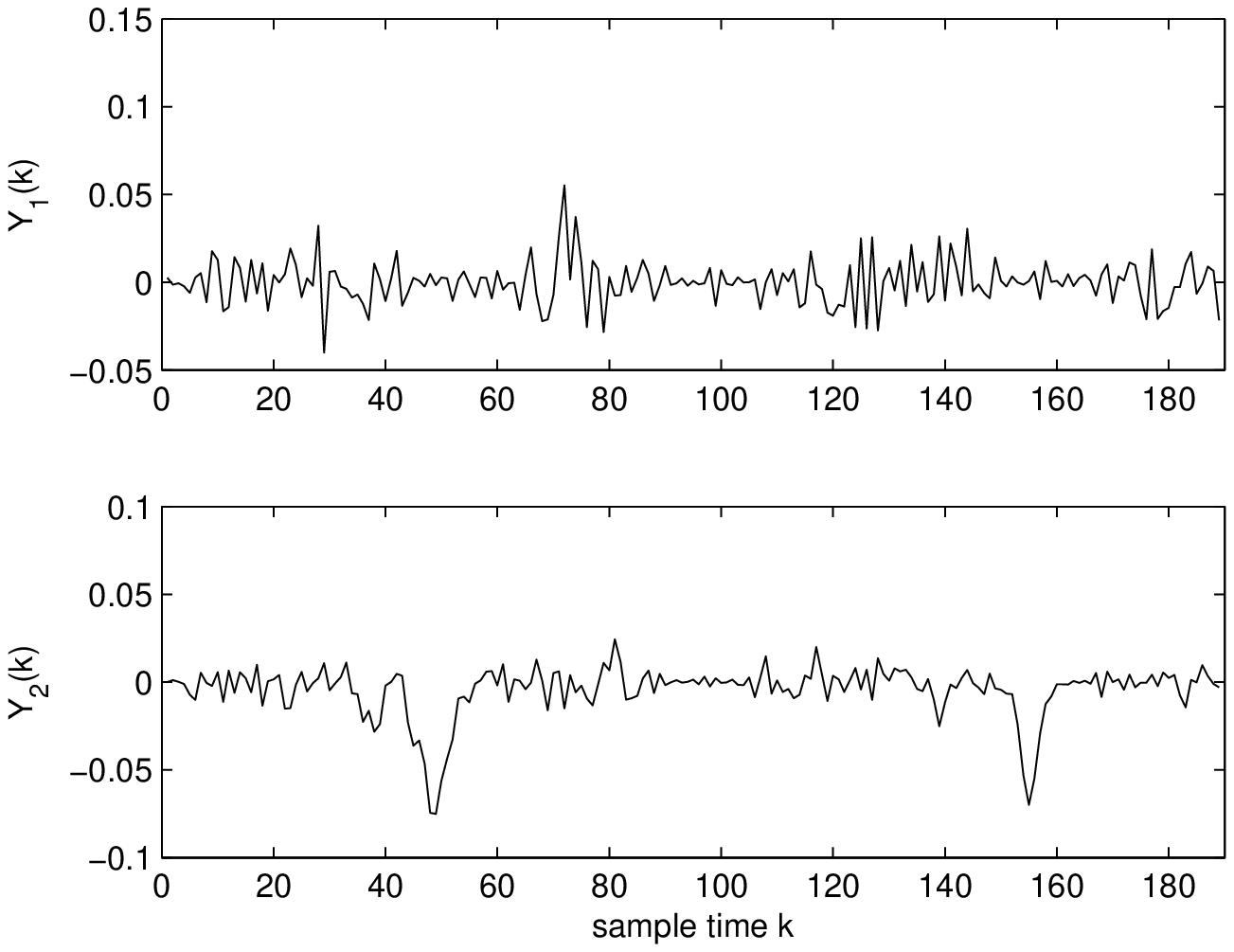}}
\caption{Training errors of the system outputs for the two trajectory tracking problems.}
\end{figure}

\subsection{``Step" Trajectory Tracking}

\begin{figure}[!t]
\subfigure[Controlled Outputs -- ``Step"]{
\centering
\label{fig:output_step}
\includegraphics[width=\tripfigWidth\textwidth]{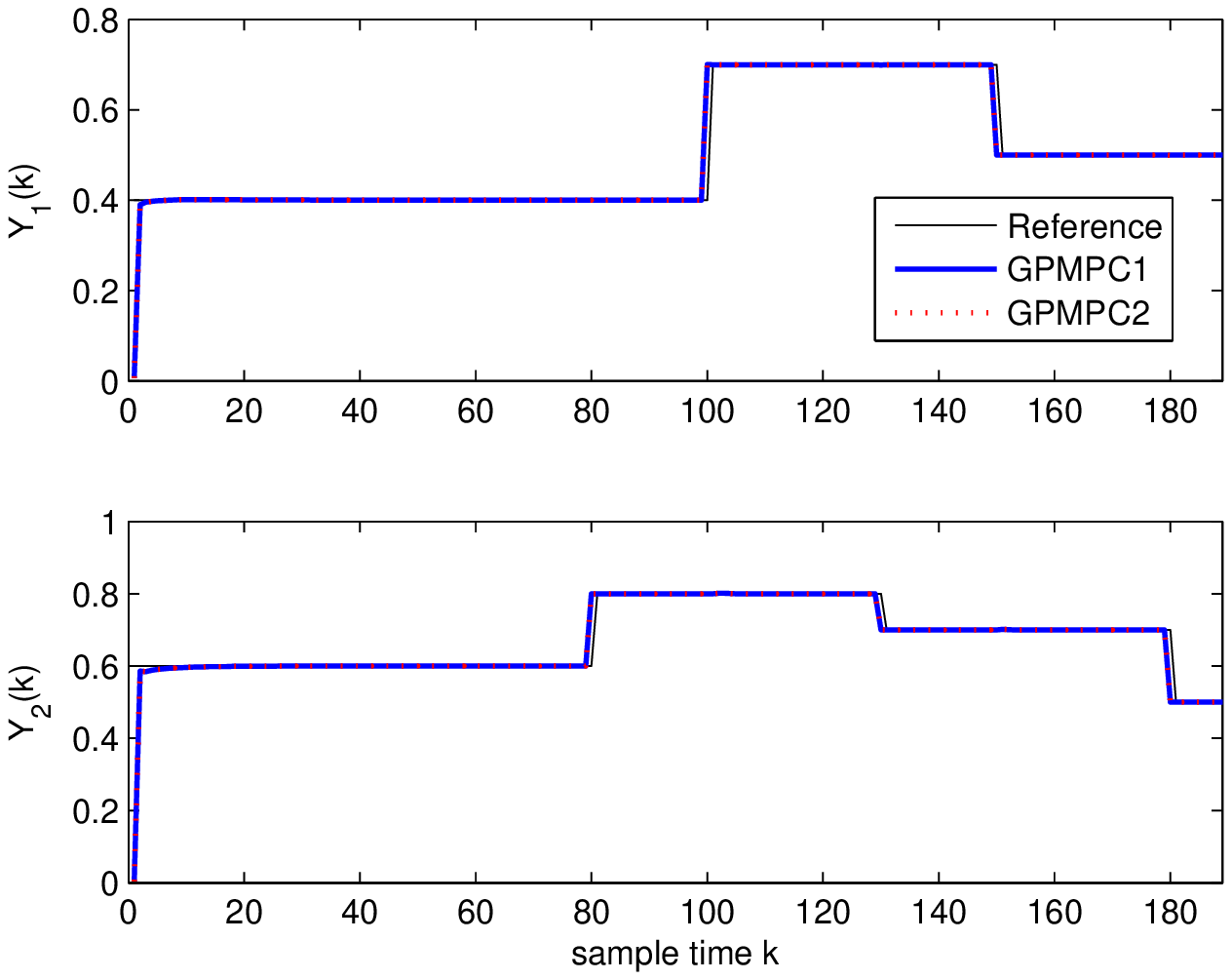}}
\subfigure[Control Inputs -- ``Step"]{
\centering
\label{fig:input_step}
\includegraphics[width=\tripfigWidth\textwidth]{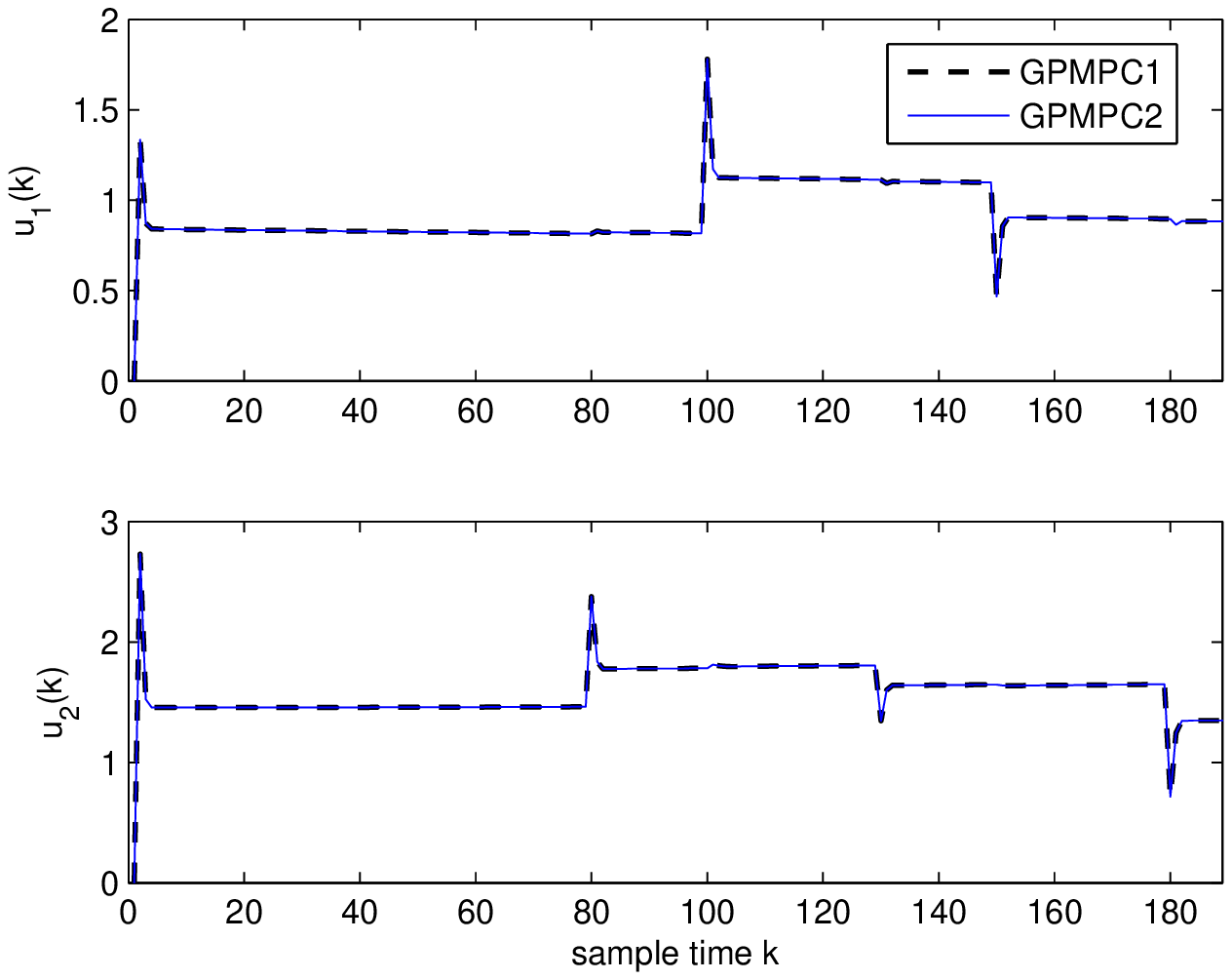}}
\subfigure[IAE -- ``Step"]{
\centering
\label{fig:iae_step}
\includegraphics[width=\tripfigWidth\textwidth]{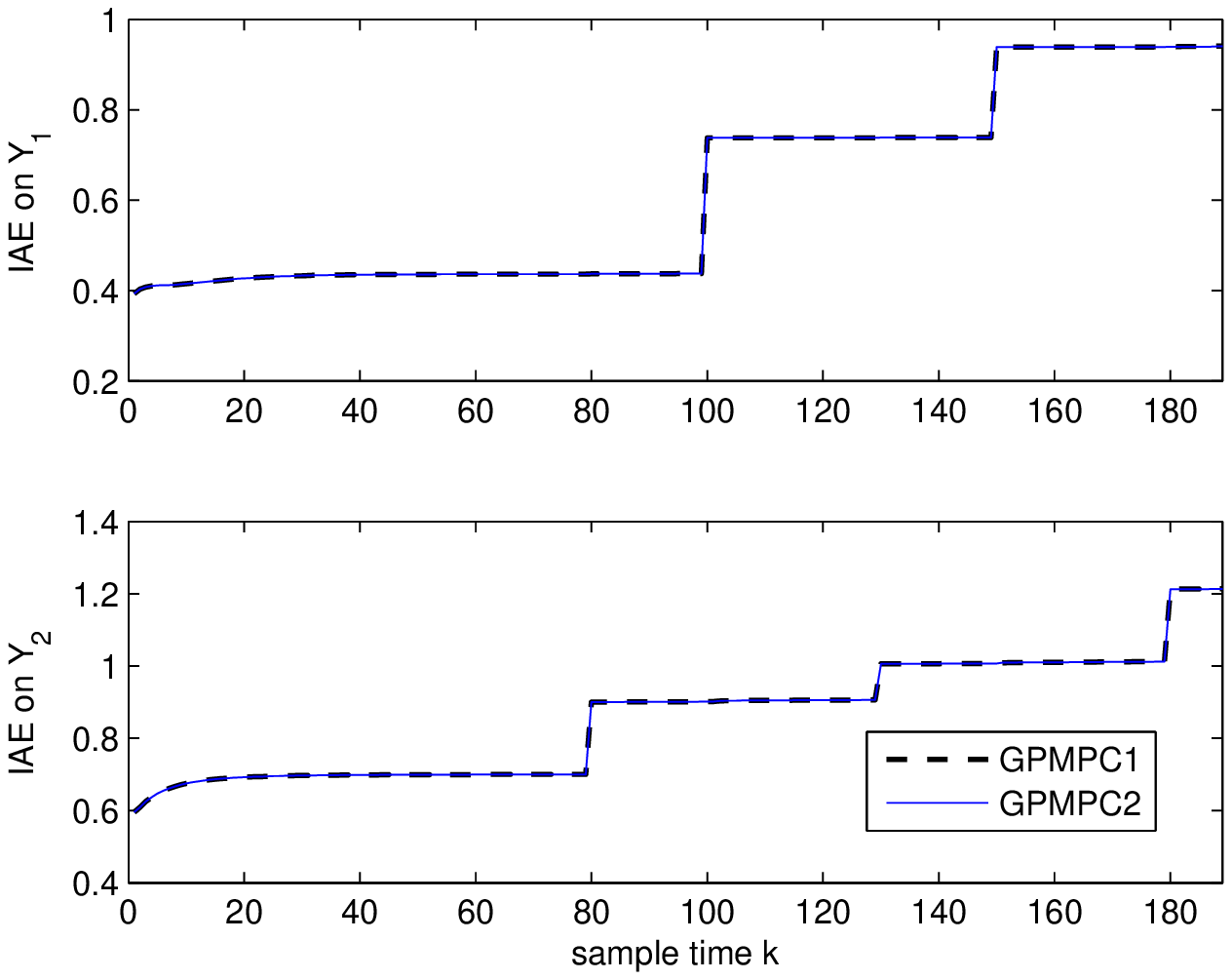}}\\
\subfigure[Controlled Outputs -- ``Lorenz"]{
\centering
\label{fig:output_lorenz_H10}
\includegraphics[width=\tripfigWidth\textwidth]{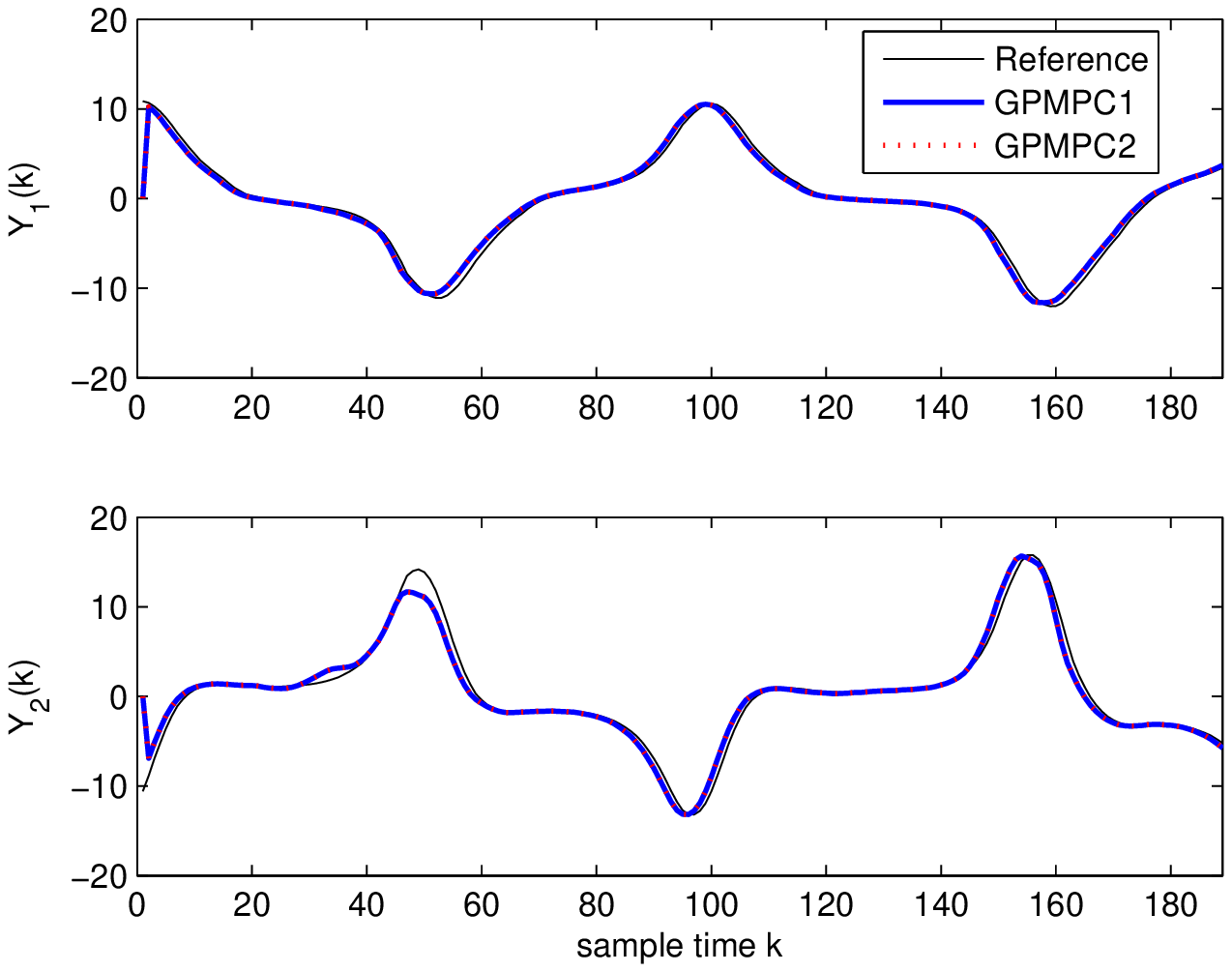}}
\subfigure[Control Inputs -- ``Lorenz"]{
\centering
\label{fig:input_lorenz_H10}
\includegraphics[width=\tripfigWidth\textwidth]{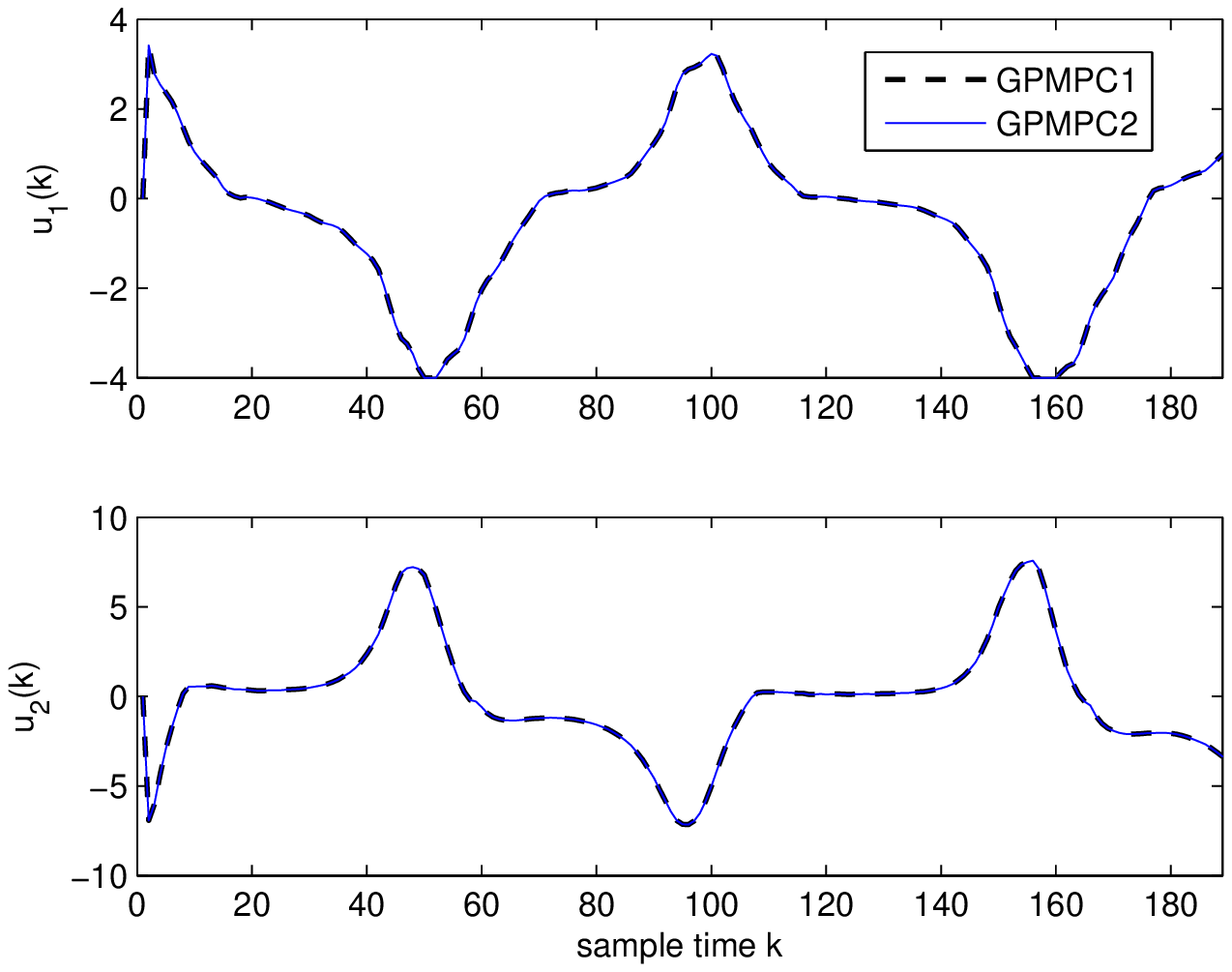}}
\subfigure[IAE -- ``Lorenz"]{
\centering
\label{fig:iae_lorenz_H10}
\includegraphics[width=\tripfigWidth\textwidth]{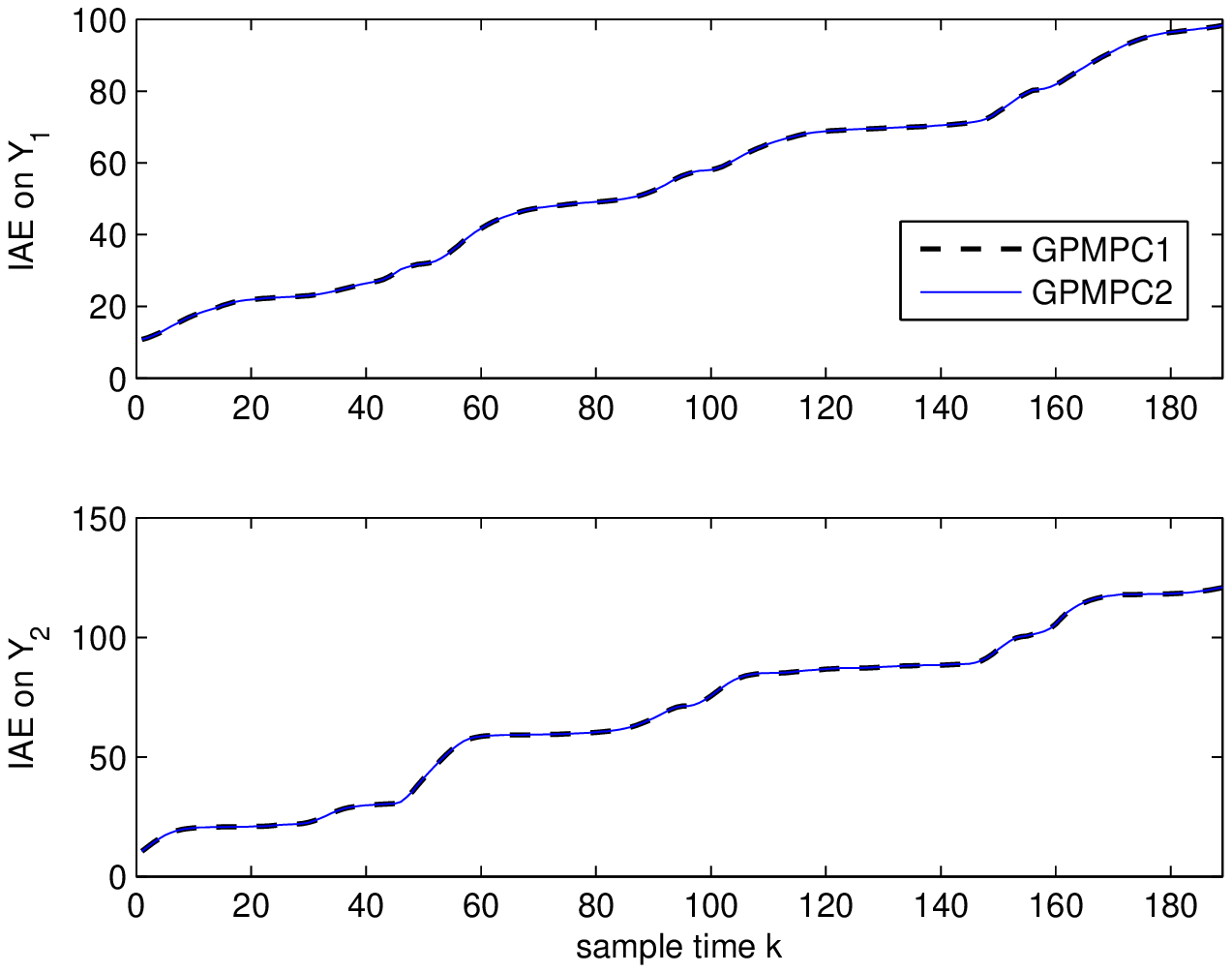}}
\caption{Simulation results of the two trajectory tracking problems.}
\end{figure}

The objective of the first experiment is to steer the nonlinear system to follow a step trajectory 
shown as the reference in Figure~\ref{fig:output_step}.
The system inputs are subjected to the following constraints:
\begin{equation}
\nonumber 0\leq u_1(k)\leq 5,\quad 0\leq u_2(k)\leq 5
\end{equation}

To generate the observations for \gls{gp} modelling,
this problem is first solved by using the \gls{nmpc} strategy proposed in~\cite{B-Grune-NMPC-2011}.
$189$ observations are collected and are used to train the~\gls{gp} models. 
The learning process took approximately $2.1$ seconds,
with a training~\gls{mse} of $9.9114\times 10^{-5}$.
Figure~\ref{fig:trainerror_step} shows the training errors for the $189$ samples.
These results show that the system is accurately learnt by using the~\gls{gp} models.

The~\gls{mpc} parameters in this simulation are:
initial states $\mathbf{x}_0=[0,0,0,0]^T$ and initial control inputs $\mathbf{u}_0=[0,0]^T$,
weighting matrix $\mathbf{Q}=\mathbf{I}_{4\times 4}$ and $\mathbf{R}=\mathbf{I}_{2\times 2}$.
In addition, the prediction horizon $H$ is $10$.
Theoretically, 
a long enough $H$ is necessary to guarantee the stability of~\gls{mpc} controllers.
However,
the complexity of~\gls{mpc} problem increases exponentially with increasing $H$.
This value of $H$ is chosen as a trade-off between the control performance and computational complexity.

The resultant controlled outputs and control inputs by using GPMPC1 and GPMPC2 
are shown in Figures~\ref{fig:output_step} and \ref{fig:input_step}, respectively.
They show that both algorithms exhibit equally good control performances in this task 
since they both produced outputs close to the target.
The \gls{iae} values can be found in Figure~\ref{fig:iae_step}.

GPMPC1 takes on average $34.1$ seconds to compute the $189$ optimized control inputs.
However, GPMPC2 only requires $4.51$ seconds which is more than $8$ times more efficient than GPMPC1.
This shows the advantage in our formulation of the problem as convex optimization.

\subsection{``Lorenz" Trajectory Tracking}

\begin{figure}[!t]
	\label{fig:compare-nonlinear}
	\subfigure[$Y_1$ -- ``Lorenz"]{
		\centering
		\label{fig:methodscompare_y1}
		\includegraphics[width=\singlefigWidth\textwidth]{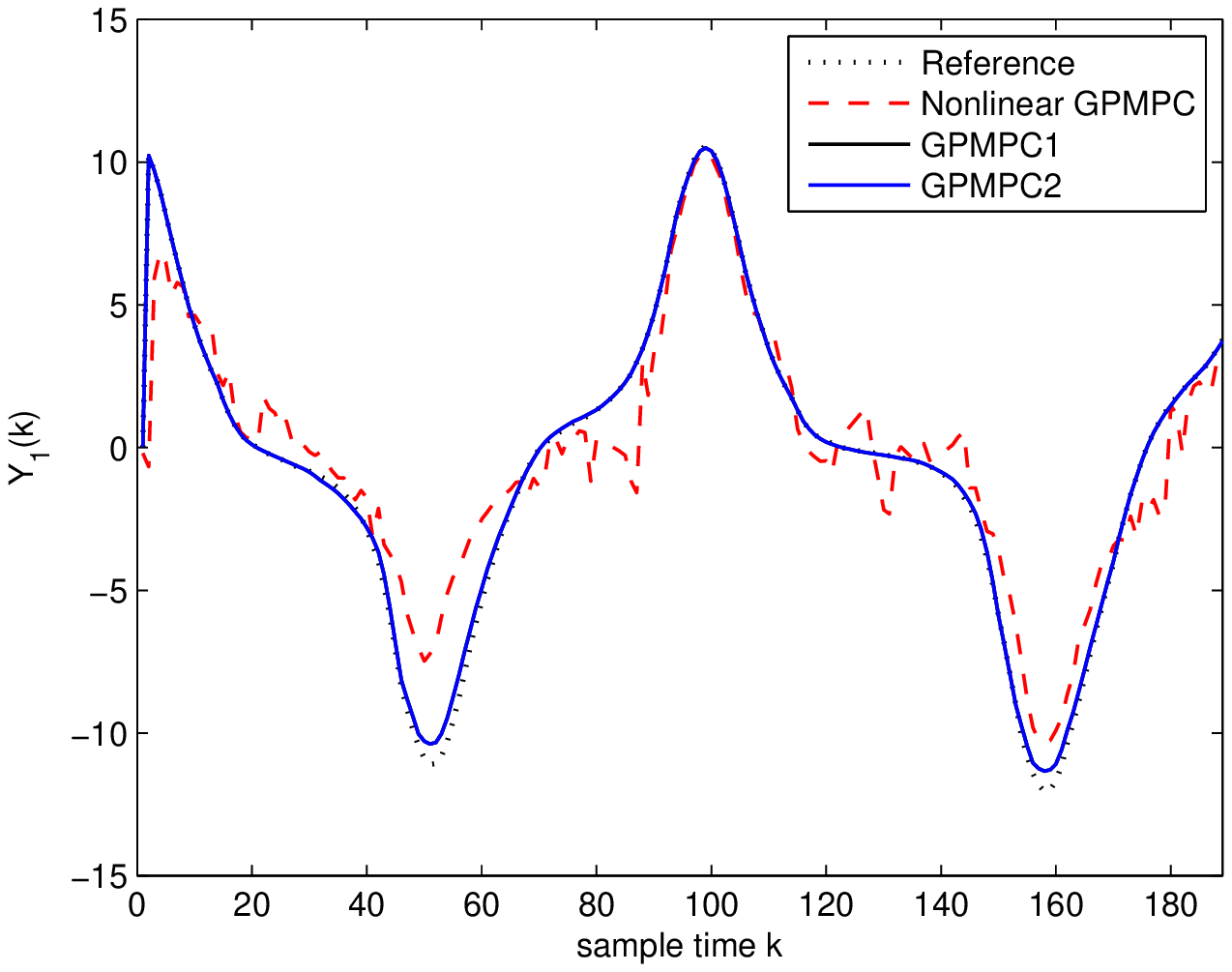}}
	\subfigure[$Y_2$ -- ``Lorenz"]{
		\label{fig:methodscompare_y2}
		\includegraphics[width=\singlefigWidth\textwidth]{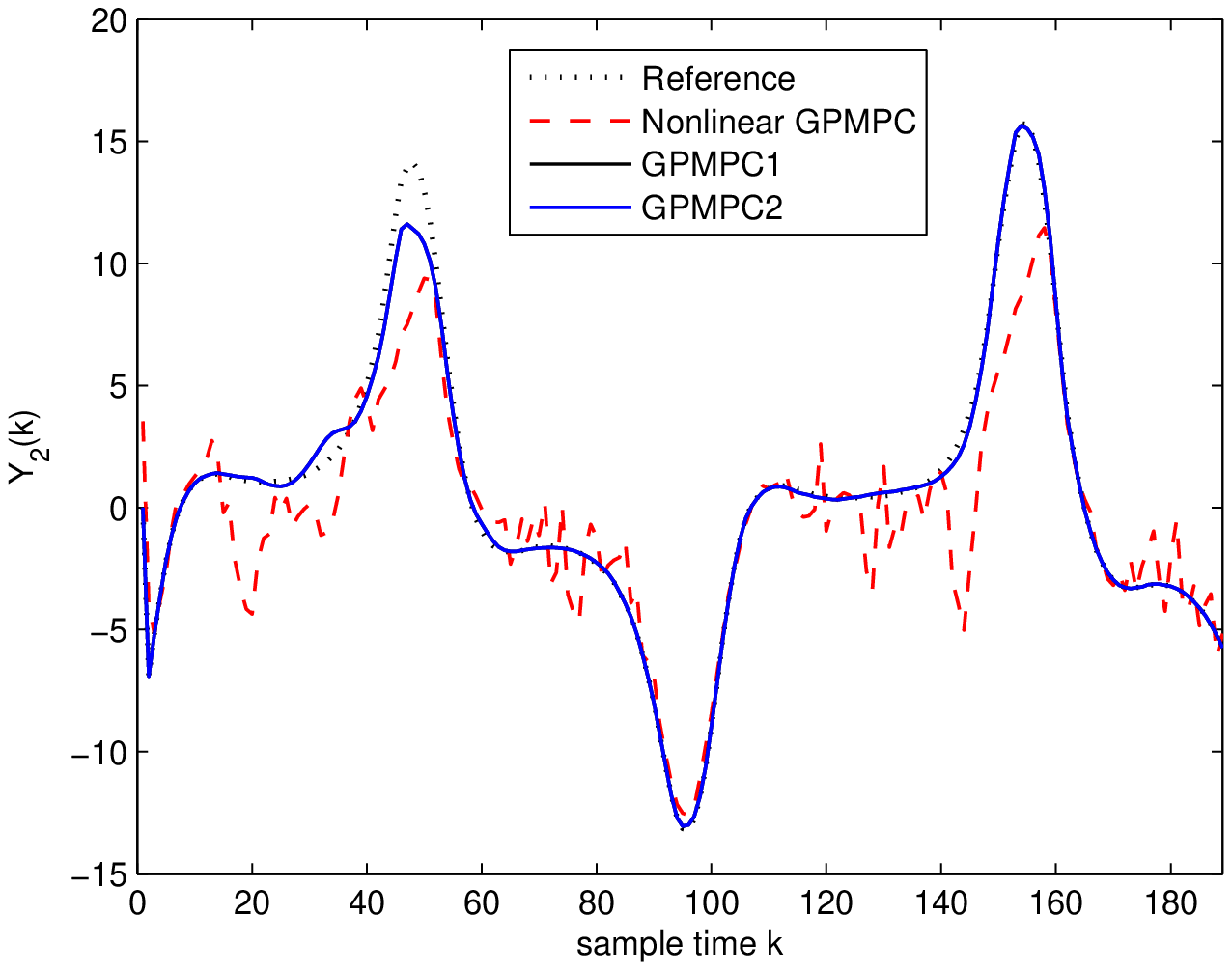}}
	\caption{Comparison of tracking performance of GPMPC1, GPMPC2 and nonlinear GPMPC in 
		\cite{IC-Kocijan-MPCusingGP-2005} for the Lorenz trajectory.}
\end{figure}

The second problem is to track a ``Lorenz" trajectory as shown in Figure~\ref{fig:output_lorenz_H10}.
In this case, the constraints on the control inputs are:
\begin{equation}
\nonumber-4\leq u_1(k)\leq 4,\quad -7\leq u_2(k)\leq 7
\end{equation}
Similar to the previous experiment,
the~\gls{nmpc} method is used to generate $189$ observations for training the \gls{gp} model.
Training time is approximately $2.4$ seconds with
a training~\gls{mse} of $0.0196$.
Figure~\ref{fig:trainerror_lorenz} shows the training error.

The~\gls{mpc} parameters are:
initial states $\mathbf{x}_0=[0,0,0,0]^T$,  initial control inputs $\mathbf{u}_0=[0,0]^T$,
prediction horizon $H=10$,
weighting matrix $\mathbf{Q}=\mathbf{I}_{4\times 4}$ and $\mathbf{R}=\text{diag}\{[21000, 27000]\}$.

The tracking results can be found in Figures~\ref{fig:output_lorenz_H10}, 
\ref{fig:input_lorenz_H10} and \ref{fig:iae_lorenz_H10}.
They demonstrate again that the control performance GPMPC1 and GPMPC2 are virtually the same.
In this case, on average GPMPC2 is about $5$ times more efficient than GPMPC1 ($5.38$ seconds versus $24.72$ seconds).

The performance of the two proposed algorithms is compared with the 
	nonlinear GPMPC proposed in \cite{IC-Kocijan-MPCusingGP-2005}.
	Even though problem (\ref{eqn:optproblem2}) with cost function (\ref{eqn:costfunction2-2}) 
	is more complicated than the one considered in \cite{IC-Kocijan-MPCusingGP-2005}, they are essentially similar.
	Tracking results for $H=1$ are shown in Figure~\ref{fig:compare-nonlinear}.
	 show that the both two proposed algorithms outperform than the nonlinear GPMPC in the ``Lorenz" trajectory tracking problem. 
	In addition, the GPMPC1 and GPMPC2 only require approximately $5$ and $7$ seconds to compute all $189$ control actions, compared to $150$ seconds used in nonlinear GPMPC. 

\subsection{Sensitivity to Training Data}

\begin{table}[!t]
	\setlength{\tabcolsep}{12pt}
	\caption{MSE values for Lorenz trajectory tracking problem with GPMPC1 and GPMPC2 using
		GP models with different amount of training data.}
	\label{tb:learningresults}
	\centering
	\begin{tabular}{c| c| c| c| c| c| c}
		\hline
		\multirow{2}{*}{} &\multicolumn{3}{c|}{Model for GPMPC1} &\multicolumn{3}{c}{Model for GPMPC2}\\
		\hline
		Training Data& $60\%$ & $80\%$ & $100\%$ & $60\%$ & $80\%$ & $100\%$\\
		\hline
		$Y_1$ & $4.7831$ & $1.36$ & $0.0528$ & $4.6493$ & $0.6879$& $0.0539$\\
		\hline
		$Y_2$ & $10.7518$ & $1.0960$ & $0.2995$ & $6.6748$ & $2.1522$& $0.3085$\\
		\hline
	\end{tabular}
\end{table}

\begin{figure}[!t]
	\subfigure[$Y_1$ -- ``Lorenz"]{
		\centering
		\label{fig:modelscompare_y1}
		\includegraphics[width=\singlefigWidth\textwidth]{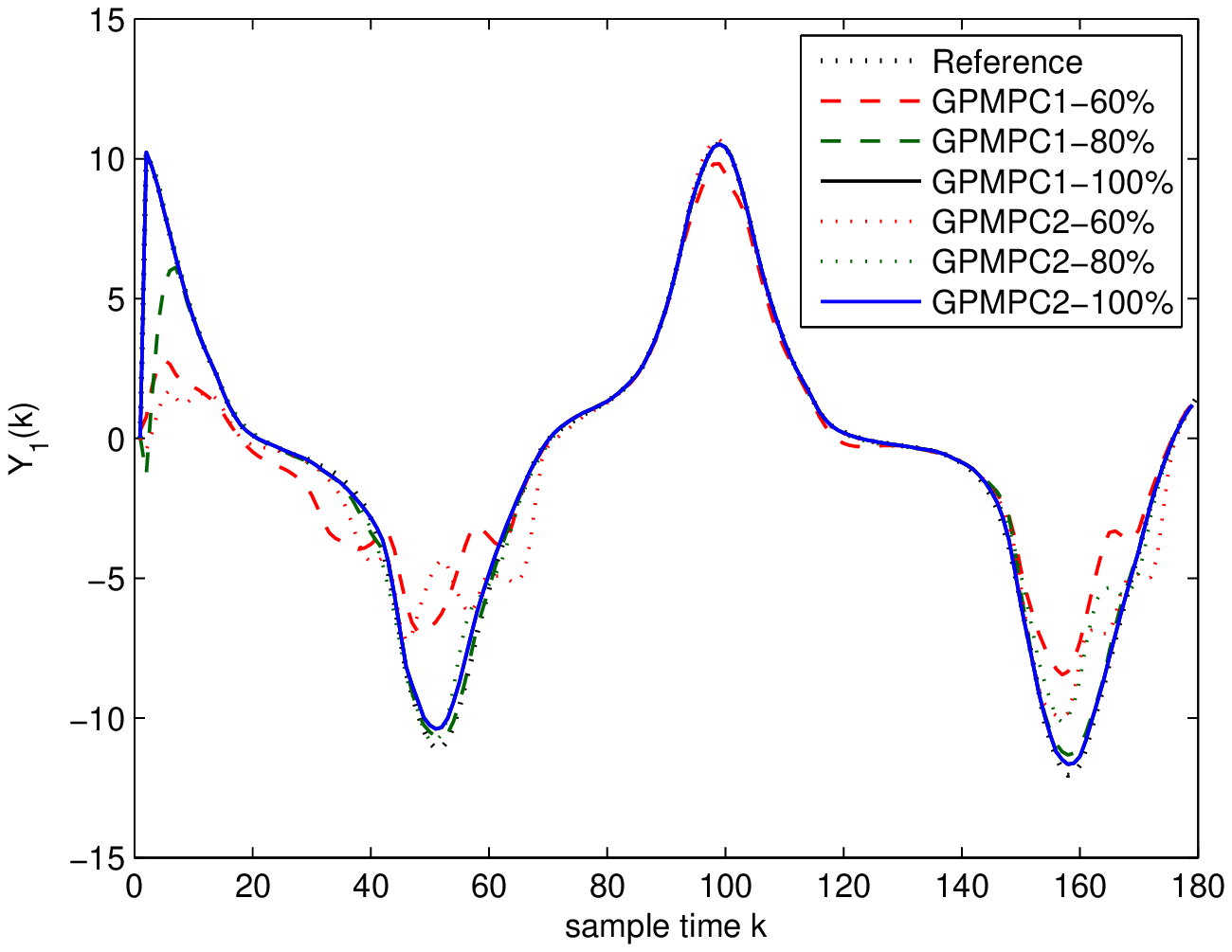}}
	\subfigure[$Y_2$ -- ``Lorenz"]{
		\label{fig:modelscompare_y2}
		\includegraphics[width=\singlefigWidth\textwidth]{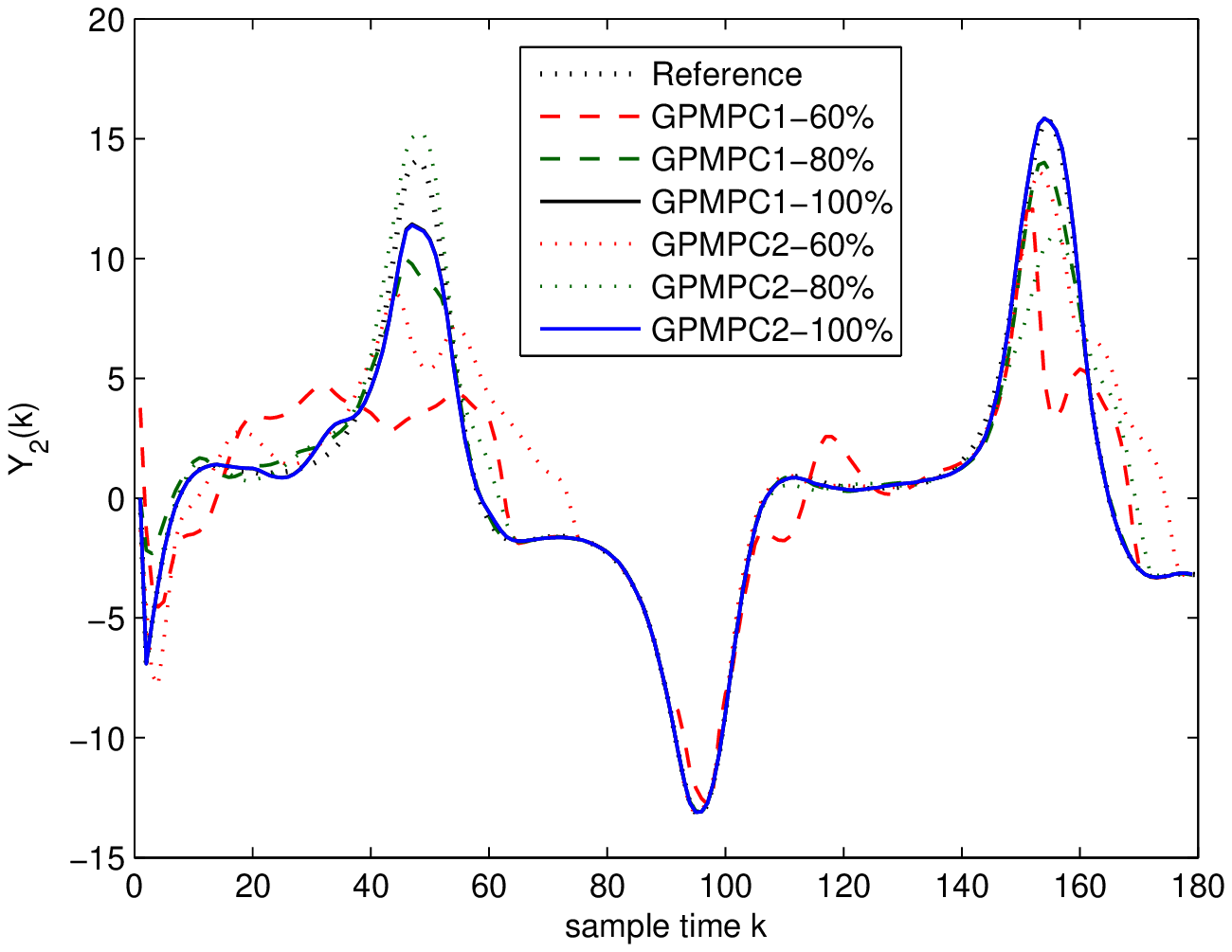}}
	\caption{Comparisons between proposed approaches with different learnt GP models in the ``Lorenz" trajectory tracking problem.}
	\label{fig:modelscompare}
\end{figure}

Since the closed-loop stability of proposed GPMPC1 and GPMPC2 
	are not guaranteed as discussed in Section~\ref{subsec:stability}, it is necessary to test them with different models.
	Here, both GPMPC1 and GPMPC2 are each tested with three separate \gls{gp} models
	for the Lorenz trajectory tracking problem. 
	These models are trained by using $60\%$, $80\%$ and $100\%$ of 
	all of $179$ observations respectively.
	Figure~\ref{fig:modelscompare} shows how well each model track the reference outputs.
	Table~\ref{tb:learningresults} shows the tracking \gls{mse} values.
	These results indicate that while the models trained with $100\%$ and $80\%$ observations perform quite well,
	the ones trained with $60\%$ data are inadequate.

\section{Conclusions}
\label{sec:conclude}

Two~\gls{gp} based~\gls{mpc} approaches (GPMPC1 and GPMPC2) have been presented
for the trajectory tracking problem of an unknown nonlinear dynamical system.
The system is modelled using \gls{gp} techniques offline. 
These two approaches handle the model uncertainties in the form of \gls{gp} variances in different ways.
GPMPC1 formulated the \gls{mpc} optimization problem in such a way that
model uncertainties are treated as the slack variables of~\gls{gp} mean constraints
and are included in the objective function as the penalty term.
The resulting \gls{smpc} problem is relaxed to a deterministic 
non-convex nonlinear optimization problem.
The solution of the resultant problem is obtained using the~\gls{fpsqp} method 
based on a linearized \gls{gp} local model.
With GPMPC2, the variance forms part of the state vector.
This allows model uncertainties to be directly included in the computation of the optimized controls.
By using the extended linearized \gls{gp} local model,
the non-convex optimization problem is relaxed to a convex one which
is solved using an active-set method.
Simulation results on two different trajectories show that
both approaches perform equally well.
However, GPMPC2 is several times more efficient computationally compared with GPMPC1, 
especially for a longer horizon. 
A brief discussion on how closed-loop stability could be guaranteed reveals that
the resulting optimization problem will be different from the one considered in this paper.
This issue will be addressed in future work.

%\bibliographystyle{IEEEtran}
%\bibliography{IEEEfull,myBibDataBase}

\begin{thebibliography}{1}

	\bibitem{A-Mayne-ConstrainedMPC-StabilityOptimality-2000}
	D.~Q. Mayne, J.~B. Rawlings, C.~V. Rao, and P.~O. Scokaert, ``Constrained model
	predictive control: {S}tability and optimality,'' \emph{Automatica}, vol.~36,
	no.~6, pp. 789--814, 2000.
	
	\bibitem{A-Qin-SurveyMPC-2003}
	S.~J. Qin and T.~A. Badgwell, ``A survey of industrial model predictive control
	technology,'' \emph{Control Engineering Practice}, vol.~11, no.~7, pp.
	733--764, 2003.
	
	\bibitem{A-Mayne-ServeyMPC-2014}
	D.~Q. Mayne, ``Model predictive control: {R}ecent developments and future
	promise,'' \emph{Automatica}, vol.~50, no.~12, pp. 2967--2986, 2014.
	
	\bibitem{A-Dimitri-SurveyDataDrivenModel-2008}
	D.~P. Solomatine and A.~Ostfeld, ``Data-driven modelling: some past experiences
	and new approaches,'' \emph{Journal of hydroinformatics}, vol.~10, no.~1, pp.
	3--22, 2008.
	
	\bibitem{B-Nelles-NonlinearSystemIdentify-2013}
	O.~Nelles, \emph{Nonlinear system identification: {f}rom classical approaches
		to neural networks and fuzzy models}.\hskip 1em plus 0.5em minus 0.4em\relax
	Springer Science \& Business Media, 2013.
	
	\bibitem{A-Alamo-OpenLoopMinMaxMPC-2005}
	T.~Alamo, D.~M. de~La~Pe{\~n}a, D.~Lim{\'o}n, and E.~F. Camacho, ``Constrained
	min-max predictive control: {M}odifications of the objective function leading
	to polynomial complexity,'' \emph{{IEEE} Transactions on Automatic Control},
	vol.~50, no.~5, pp. 710--714, 2005.
	
	\bibitem{A-Limon-ISSMinMaxMPC-2006}
	D.~Lim{\'o}n, T.~Alamo, F.~Salas, and E.~F. Camacho, ``Input to state stability
	of min--max {MPC} controllers for nonlinear systems with bounded
	uncertainties,'' \emph{Automatica}, vol.~42, no.~5, pp. 797--803, 2006.
	
	\bibitem{A-Langson-TubeMPC-2004}
	W.~Langson, I.~Chryssochoos, S.~Rakovi{\'c}, and D.~Q. Mayne, ``Robust model
	predictive control using tubes,'' \emph{Automatica}, vol.~40, no.~1, pp.
	125--133, 2004.
	
	\bibitem{A-Zhang-SwitchedMPC-2016}
	L.~Zhang, S.~Zhuang, and R.~D. Braatz, ``Switched model predictive control of
	switched linear systems: {F}easibility, stability and robustness,''
	\emph{Automatica}, vol.~67, pp. 8--21, 2016.
	
	\bibitem{A-Schwarm-ChanceConstrainedMPC-1999}
	A.~T. Schwarm and M.~Nikolaou, ``Chance-constrained model predictive control,''
	\emph{American Institute of Chemical Engineers}, vol.~45, no.~8, pp.
	1743--1752, 1999.
	
	\bibitem{IP-Bernardini-ScenarioSMPC-2009}
	D.~Bernardini and A.~Bemporad, ``Scenario-based model predictive control of
	stochastic constrained linear systems,'' in \emph{{IEEE} Proceedings of
		International Conference on Decision and Control}.\hskip 1em plus 0.5em minus
	0.4em\relax IEEE, 2009, pp. 6333--6338.
	
	\bibitem{A-Cannon-ProbConstrainsSMPC-2011}
	M.~Cannon, B.~Kouvaritakis, S.~V. Rakovi{\'c}, and Q.~Cheng, ``Stochastic tubes
	in model predictive control with probabilistic constraints,'' \emph{{IEEE}
		Transactions on Automatic Control}, vol.~56, no.~1, pp. 194--200, 2011.
	
	\bibitem{IP-Mesbah-StochasticMPC-2014}
	A.~Mesbah, S.~Streif, R.~Findeisen, and R.~Braatz, ``Stochastic nonlinear model
	predictive control with probabilistic constraints,'' in \emph{American
		Control Conference}.\hskip 1em plus 0.5em minus 0.4em\relax IEEE, 2014, pp.
	2413--2419.
	
	\bibitem{IP-Fagiano-NMPCusingPC-2012}
	L.~Fagiano and M.~Khammash, ``Nonlinear stochastic model predictive control via
	regularized polynomial chaos expansions,'' in \emph{{IEEE} Proceedings of
		International Conference on Decision and Control}.\hskip 1em plus 0.5em minus
	0.4em\relax IEEE, 2012, pp. 142--147.
	
	\bibitem{B-GPMLbook-2006}
	C.~Rasmussen and C.~Williams, \emph{{Gaussian Processes
			for Machine Learning}}.\hskip 1em plus 0.5em minus 0.4em\relax Cambridge, MA,
	USA: MIT Press, 1 2006.
	
	\bibitem{IP-Zhu-Particle-2010}
	F.~Zhu, C.~Xu, and G.~Dui, ``Particle swarm hybridize with {G}aussian process
	regression for displacement prediction,'' in \emph{{IEEE} Proceedings of
		International Conference on Bio-Inspired Computing: Theories and
		Applications}.\hskip 1em plus 0.5em minus 0.4em\relax IEEE, 2010, pp.
	522--525.
	
	\bibitem{IP-Petelin-EvolvingGP-2011}
	D.~Petelin and J.~Kocijan, ``Control system with evolving {G}aussian process
	models,'' in \emph{{IEEE} Workshop on Evolving and Adaptive Intelligent
		Systems (EAIS)}.\hskip 1em plus 0.5em minus 0.4em\relax IEEE, 2011, pp.
	178--184.
	
	\bibitem{IP-Gang-2014a}
	G.~Cao, E.~M.-K. Lai, and F.~Alam, ``Particle swarm optimization for convolved
	{G}aussian process models,'' in \emph{International Joint Conference on
		Neural Networks (IJCNN)}.\hskip 1em plus 0.5em minus 0.4em\relax IEEE, 6-11
	July 2014, pp. 1573--1578.
	
	\bibitem{IP-Kocijan-MPC-GP-2004}
	J.~Kocijan, R.~Murray-Smith, C.~E. Rasmussen, and A.~Girard, ``Gaussian process
	model based predictive control,'' in \emph{American Control Conference},
	vol.~3.\hskip 1em plus 0.5em minus 0.4em\relax IEEE, 2004, pp. 2214--2219.
	
	\bibitem{IP-Grancharova-ExplicitMPC-2007}
	A.~Grancharova, J.~Kocijan, and T.~A. Johansen, ``Explicit stochastic nonlinear
	predictive control based on {G}aussian process models,'' in \emph{European
		Control Conference}, 2007, pp. 2340--2347.
	
	\bibitem{A-Klenske-GPMPC4Periodic-2015}
	E.~D. Klenske, M.~N. Zeilinger, B.~Scholkopf, and P.~Hennig, ``Gaussian
	process-based predictive control for periodic error correction,''
	\emph{{IEEE} Transactions on Control Systems Technology}, 2015.
	
	\bibitem{IP-GPMPC4LTV-Gang-2016a}
	G.~Cao, E.~M.-K. Lai, and F.~Alam, ``Gaussian process based model predictive
	control for linear time varying systems,'' in \emph{International Workshop on
		Advanced Motion Control (AMC Workshop)}.\hskip 1em plus 0.5em minus
	0.4em\relax IEEE, 22-24 April 2016.
	
	\bibitem{IP-GPMPC4Quad-Gang-2016b}
	------, ``Gaussian process model predictive control of {U}nmanned
	{Q}uadrotors,'' in \emph{International Conference on Control, Automation and
		Robotics (ICCAR)}.\hskip 1em plus 0.5em minus 0.4em\relax IEEE, 28-30 April
	2016.
	
	\bibitem{PHD-Deisenroth-EfficientRLusingGP-2010}
	M.~P. Deisenroth, ``Efficient reinforcement learning using {G}aussian
	processes,'' Ph.D. dissertation, Karlsruhe Institute of Technology, 2010.
	
	\bibitem{IP-Girard-MultiStepTimeSeriesForecasting-2003}
	A.~Girard, C.~E. Rasmussen, J.~Q. Candela, and R.~Murray-Smith, ``Gaussian
	process priors with uncertain input -- {A}pplication to multiple-step ahead
	time series forecasting,'' in \emph{Advances in Neural Information Processing
		Systems}.\hskip 1em plus 0.5em minus 0.4em\relax MIT, 2003, pp. 545--552.
	
	\bibitem{IP-Candela-GPUncertainPropagation-2003}
	J.~Q. Candela, A.~Girard, J.~Larsen, and C.~E. Rasmussen, ``Propagation of
	uncertainty in bayesian kernel models-application to multiple-step ahead
	forecasting,'' in \emph{{IEEE} Proceedings of International Conference on
		Acoustics, Speech, and Signal Processing (ICASSP)}, vol.~2.\hskip 1em plus
	0.5em minus 0.4em\relax IEEE, 2003, pp. II--701.
	
	\bibitem{A-Sparse-Appro-2005}
	J.~Qui{\~n}onero-Candela and C.~E. Rasmussen, ``A unifying view of sparse
	approximate {G}aussian process regression,'' \emph{Journal of Machine
		Learning Research}, vol.~6, pp. 1939--1959, 2005.
	
	\bibitem{IP-Berkenkamp-RboustLBNMPC-2014}
	F.~Berkenkamp and A.~P. Schoellig, ``Learning-based robust control:
	{G}uaranteeing stability while improving performance,'' in \emph{{IEEE/RSJ}
		Proceedings of International Conference on Intelligent Robots and Systems
		(IROS)}, 2014.
	
	\bibitem{IP-Pan-PDDP-2014}
	Y.~Pan and E.~Theodorou, ``Probabilistic differential dynamic programming,'' in
	\emph{Advances in Neural Information Processing Systems}, 2014, pp.
	1907--1915.
	
	\bibitem{A-Grancharova-ExplicitMPC-2008}
	A.~Grancharova, J.~Kocijan, and T.~A. Johansen, ``Explicit stochastic
	predictive control of combustion plants based on {G}aussian process models,''
	\emph{Automatica}, vol.~44, no.~6, pp. 1621--1631, 2008.
	
	\bibitem{A-Troltzsch-LagranMultipler-2005}
	F.~Tr{\"o}ltzsch, ``Regular {L}agrange multipliers for control problems with
	mixed pointwise control-state constraints,'' \emph{SIAM Journal on
		Optimization}, vol.~15, no.~2, pp. 616--634, 2005.
	
	\bibitem{IP-Diehl-EfficientSolution4MPC-2009}
	M.~Diehl, H.~J. Ferreau, and N.~Haverbeke, ``Efficient numerical methods for
	nonlinear {MPC} and moving horizon estimation,'' in \emph{International
		Workshop on assessment and future directions on Nonlinear Model Predictive
		Control}.\hskip 1em plus 0.5em minus 0.4em\relax Pavia, Italy: Springer,
	2008, pp. 391--417.
	
	\bibitem{A-Lucidi-DeviativeFreeOpt-2002}
	S.~Lucidi, M.~Sciandrone, and P.~Tseng, ``Objective-derivative-free methods for
	constrained optimization,'' \emph{Mathematical Programming}, vol.~92, no.~1,
	pp. 37--59, 2002.
	
	\bibitem{A-Liuzzi-SequentialDerivativeFreeOpt-2010}
	G.~Liuzzi, S.~Lucidi, and M.~Sciandrone, ``Sequential penalty derivative-free
	methods for nonlinear constrained optimization,'' \emph{SIAM Journal on
		Optimization}, vol.~20, no.~5, pp. 2614--2635, 2010.
	
	\bibitem{A-Luo-PSO4NLP-2007}
	L.~Yiqing, Y.~Xigang, and L.~Yongjian, ``An improved {PSO} algorithm for
	solving non-convex {NLP/MINLP} problems with equality constraints,''
	\emph{Computers \& chemical engineering}, vol.~31, no.~3, pp. 153--162, 2007.
	
	\bibitem{A-Yeniay-GA4NLP-2005}
	O.~Yeniay, ``Penalty function methods for constrained optimization with genetic
	algorithms,'' \emph{Mathematical and Computational Applications}, vol.~10,
	no.~1, pp. 45--56, 2005.
	
	\bibitem{A-Wright-FTRSQP-2004}
	S.~J. Wright and M.~J. Tenny, ``A feasible trust-region sequential quadratic
	programming algorithm,'' \emph{SIAM journal on optimization}, vol.~14, no.~4,
	pp. 1074--1105, 2004.
	
	\bibitem{A-Peng-FeasibleTRSQP-2006}
	Y.-h. Peng and S.~Yao, ``A feasible trust-region algorithm for inequality
	constrained optimization,'' \emph{Applied mathematics and computation}, vol.
	173, no.~1, pp. 513--522, 2006.
	
	\bibitem{A-Zhang-AdaptiveTrustRegion-2002}
	X.~Zhang, J.~Zhang, and L.~Liao, ``An adaptive trust region method and its
	convergence,'' \emph{Science in China Series A: Mathematics}, vol.~45, no.~5,
	pp. 620--631, 2002.
	
	\bibitem{A-Tenny-FPSQP4NLP-2004}
	M.~J. Tenny, S.~J. Wright, and J.~B. Rawlings, ``Nonlinear model predictive
	control via feasibility-perturbed sequential quadratic programming,''
	\emph{Computational Optimization and Applications}, vol.~28, no.~1, pp.
	87--121, 2004.
	
	\bibitem{A-Bemporad-ExplicitLinearQuadRegulorMPC-2002}
	A.~Bemporad, M.~Morari, V.~Dua, and E.~N. Pistikopoulos, ``The explicit linear
	quadratic regulator for constrained systems,'' \emph{Automatica}, vol.~38,
	no.~1, pp. 3--20, 2002.
	
	\bibitem{A-Wang-FastMPC-2010}
	Y.~Wang and S.~Boyd, ``Fast model predictive control using online
	optimization,'' \emph{{IEEE} Transactions on Control Systems Technology},
	vol.~18, no.~2, pp. 267--278, 2010.
	
	\bibitem{B-Fletcher-PracticalMethods4Opt-1987}
	R.~Fletcher, \emph{Practical methods of optimization}, 2nd~ed.\hskip 1em plus
	0.5em minus 0.4em\relax Wiley-Interscience Publication, 1987.
	
	\bibitem{A-Pan-MPCusingANN-2012}
	Y.~Pan and J.~Wang, ``Model predictive control of unknown nonlinear dynamical
	systems based on recurrent neural networks,'' \emph{{IEEE} Transactions on
		Industrial Electronics}, vol.~59, no.~8, pp. 3089--3101, 2012.
	
	\bibitem{B-Grune-NMPC-2011}
	L.~Gr{\"u}ne and J.~Pannek, \emph{Nonlinear model predictive control--Theory
		and Algorithms.}\hskip 1em plus 0.5em minus 0.4em\relax London, U.K:
	Springer-Verlag, 2011.
	
	\bibitem{IC-Kocijan-MPCusingGP-2005}
	J.~Kocijan and R.~Murray-Smith, ``Nonlinear predictive control with a
	{G}aussian process model,'' in \emph{In R. Murray-Smith and R. Shorten
		(eds.), Switching and Learning in Feedback Systems}.\hskip 1em plus 0.5em
	minus 0.4em\relax Springer, 2005, pp. 185--200.
	
\end{thebibliography}

% Generated by IEEEtran.bst, version: 1.14 (2015/08/26)

\end{document}